\documentclass[a4paper,fleqn,usenatbib]{mnras}

\usepackage{newtxtext,newtxmath}
\usepackage{mathptmx}

\usepackage[T1]{fontenc}
\usepackage{ae,aecompl}

\usepackage{graphicx}	
\usepackage{amsmath}	
\usepackage{amssymb}	
\usepackage{epsfig}
\usepackage{blindtext, subfig}

\title{H$_{2}$ emission from non-stationary magnetized bow shocks}

\author[Tram, Lesaffre et al]{
L.N.Tram$^{1,4}$\thanks{E-mail: le.ngoctram@lra.ens.fr},
P.Lesaffre$^{1}$,
S.Cabrit$^{2}$,
A.Gusdorf$^{1}$, 
and P.T.Nhung$^{3}$
\\
$^{1}$LERMA, Observatoire de Paris, \'Ecole Normal Sup\'erieure, PSL Research University, CNRS, Sorbonne Universit\'es,\\ UPMC Univ. Paris 06, F-75231, Paris, France\\
$^{2}$LERMA, Observatoire de Paris, PSL Research University, CNRS, SorbonneUniversit\'es, UPMC Univ. Paris 06,\\ F-75014, Paris, France\\
$^{3}$ Department of AstroPhysics (DAP) - Vietnam National Satellite Center (VNSC), Vietnam\\
$^{4}$ University of Science and Technology of Hanoi (USTH), Hanoi, Vietnam
}

\pubyear{2017}

\begin{document}
\label{firstpage}
\pagerange{\pageref{firstpage}--\pageref{lastpage}}
\maketitle

\begin{abstract}
When a fast moving star or a protostellar jet hits an interstellar cloud, the surrounding gas gets heated and illuminated: a bow shock is born which delineates the wake of the impact. In such a process, the new molecules that are formed and excited in the gas phase become accessible to observations. In this article, we revisit models of H$_{2}$ emission in these bow shocks. We approximate the bow shock by a statistical distribution of planar shocks computed with a magnetized shock model. We improve on previous works by considering arbitrary bow shapes, a finite irradiation field, and by including the age effect of non-stationary C-type shocks on the excitation diagram and line profiles of H$_{2}$. We also examine the dependence of the line profiles on the shock velocity and on the viewing angle: we suggest that spectrally resolved observations may greatly help to probe the dynamics inside the bow shock. For reasonable bow shapes, our analysis shows that low velocity shocks largely contribute to H$_{2}$ excitation diagram. This can result in an observational bias towards low velocities when planar shocks are used to interpret H$_{2}$ emission from an unresolved bow. We also report a large magnetization bias when the velocity of the planar model is set independently. Our 3D models reproduce excitation diagrams in BHR71 and Orion bow shocks better than previous 1D models. Our 3D model is also able to reproduce the shape and width of the broad H$_{2}$ 1-0S(1) line profile in an Orion bow shock.
\end{abstract}




\section{Introduction}
Jets or winds are generated in the early stages and the late phases of stellar evolution. The impact of high velocity flows on the interstellar medium (ISM) creates a shock. When the star moves with respect to the surrounding
gas, or when the tip of a jet penetrates the ISM, the shock
working surface assumes a  curved shape called 'bow shock'.

The angle between the impinging gas velocity and the
normal to the shape can vary along this bow. It also affects the angle of the ambient magnetic
field.  As a result, the local effective entrance velocity and the
transverse magnetic field change along the shock working surface. This
leads to differences in the local physical and chemical conditions,
which cause varying emission properties throughout the bow shock.

As a result, the global emission spectrum of a bow shock is expected to differ from that of a 1D plane parallel
shock. Accurate modeling of the emission properties of bow shocks is thus an important goal if we wish to retrieve essential properties of the system 
from observations, such as the propagation speed, the age,  the environment density and the magnetic field. 
Molecular hydgrogen is a particularly important tracer, as it dominates the shock cooling up to the dissociation limit (if the pre shock medium is molecular),
and it emits numerous lines from a wide range of upper energy levels within a single spectrometer setting
(in the K-band for ro-vibrational $\Delta v=1$ lines; in the mid-IR range $8-28\mu$m for the first pure rotational lines). In principle, the H$_2$ emission originating from bow shocks can be predicted by performing
2D or 3D numerical simulations, but the latter have been so far limited to single-fluid "jump" shocks, J-type (e.g. \citealt{Raga02, Suttner97}). 
Up to now they cannot treat "continuous" C-type shocks, where ion-neutral decoupling occurs in a magnetic precursor (\citealt{Draine93}). Such situation is encountered in the bow shock whenever the entrance speed drops below the magnetosonic speed in the charged fluid. To address this case, a second approach to predict H$_2$ emission from bow shocks is to prescribe a bow shape and treat each surface element as an independent 1D plane-parallel J-type or C-type shock, assuming that the emission zone remains small with respect to the local curvature. This approach was first introduced by \citet{S90} and \citet{S91} using simplified equations for the 1D shock structure and cooling. The validity of this approach was recently investigated by \citet{Kristensen08} and \citet{Gustafsson10} using refined 1D steady-state shock models that solve the full set of magneto-hydrodynamical equations with non-equilibrium chemistry, ionisation, and cooling.

\citet{Kristensen08} studied high angular resolution H$_2$ images of a bow shock in the Orion BN-KL outflow region, performing 
several 1D cuts orthogonal to the bow trace in the plane of the sky. They 
fitted each cut separately with 1D
steady shock models proposed by \citet{FP03}. They found that the resolved width 
required C-shocks, and that the variation of the fitted shock velocity and transverse magnetic field 
along the bow surface was consistent with a steady bow shock propagating into a uniform medium. This result provided some validation for the
'local 1D-shock approximation' when modeling H$_2$ emission in bow shocks, at least in this parameter regime. Following this idea,
\citet{Gustafsson10}  built 3D stationary models of bow shocks
by stitching together 1D shock models. 
They then projected their models to produce maps of the H$_2$ emission in several lines which they compared directly to
observations. They obtained better results than
\citet{Kristensen08} thanks to the ability of the 3D model to account
both for the inclination of the shock surface, with respect to the line of sight, and the multiple shocks included in the depth of their 1D cuts. 
The width of the emission maps was better reproduced. 

In this article, we 
 extend 
\citet{Gustafsson10}'s works on
H$_2$ emission 
by computing
the excitation diagram and line profiles integrated over the bow, and by considering the effect of short ages where C-shocks have not yet reached steady-state. 
Our method also increases the scope of \citet{Gustafsson10} to arbitrary bow shapes (we do not restrict the bow shape profile to power laws). 
Using time-dependent simulations, \citet{Ch1998} discovered that young C-type shocks, the age of which is smaller than the ion crossing time,
are composed of a magnetic precursor and a relaxation layer separated by an adiabatic J-type front. \citet{PL04b} later showed that the magnetic precursor and the relaxation layer were truncated stationary models of C-type and J-type shocks, respectively. 
In the present work, we make use of these CJ-type shocks to explore the age dependence of the
H$_2$ emission. Non-steady shocks are more likely to occur in low density media, where the time-scales are generally longer than those driving the mechanisms of these shocks: hence, we consider lower densities than \citet{Gustafsson10}, down to 100 cm$^{-3}$. As in \citet{PL13}, we include the grain component as part of the charged fluid, which singificantly lowers the magnetosonic speed. In addition the Paris-Durham code (\citealt{F03,FP15}), recently improved by \citet{PL13}, now allows
to consider finite UV irradiation conditions and we use a standard interstellar irradiation field of 
$G_0$=1 (\citealt{D78}) throughout the paper. This lowers slighlty further the magnetosonic speed as the ionisation degree/fraction increases but we checked it does not introduce critical changes for the H$_2$ emission properties. The lower magnetosonic speed above which no C-shock propagates and the truncated precursor in young CJ-type shocks both act in a way so that they give more weight to J-type shocks compared to \citet{Gustafsson10}, who had their J-type shocks H$_2$ emission dimmed by dissociation above $\simeq$15-20 km s$^{-1}$ due to the larger densities. Finally, we also investigate the line profiles which were not examined by \citet{Gustafsson10}.
 
We study how the geometry influences the distribution of shock
entrance velocity and transverse magnetic field in section
\ref{sec:model}.
We present our grid of planar shock models at finite
ages in section \ref{sec:1D_shock}. In the next section 4, we combine the planar shock models to build 3D models of bow shocks. We examine the observable H$_2$ excitation diagram and the potential biases which arise when
1D models are fit to intrinsically 3D models. We apply our 3D model to constrain parameters of the BHR71 bipolar outflow and for a bow shock in Orion. Finally, We study the properties of H$_{2}$ line profiles and show how it can be used to retrieve dynamical information. We summarise and conclude in section 5. 

\section{The model}
\label{sec:model}

As in \cite{Gustafsson10}, we assume that the 3D bow shock is made of independent planar shocks. 
In fact, we neglect the curvature effects and the friction between different 1D shock layers, the gradients of entrance conditions in the planar shock models, and the possible geometrical dilatation in the post-shock: 
our approximation is valid as long as the curvature radius of the bow shock is large with respect to the emitting thickness
of the working surface.

\subsection{Geometry and coordinate system} \label{sec:geometry_shock}
\begin{figure}
\raggedright
 		\includegraphics[width=1.165\linewidth]
 		{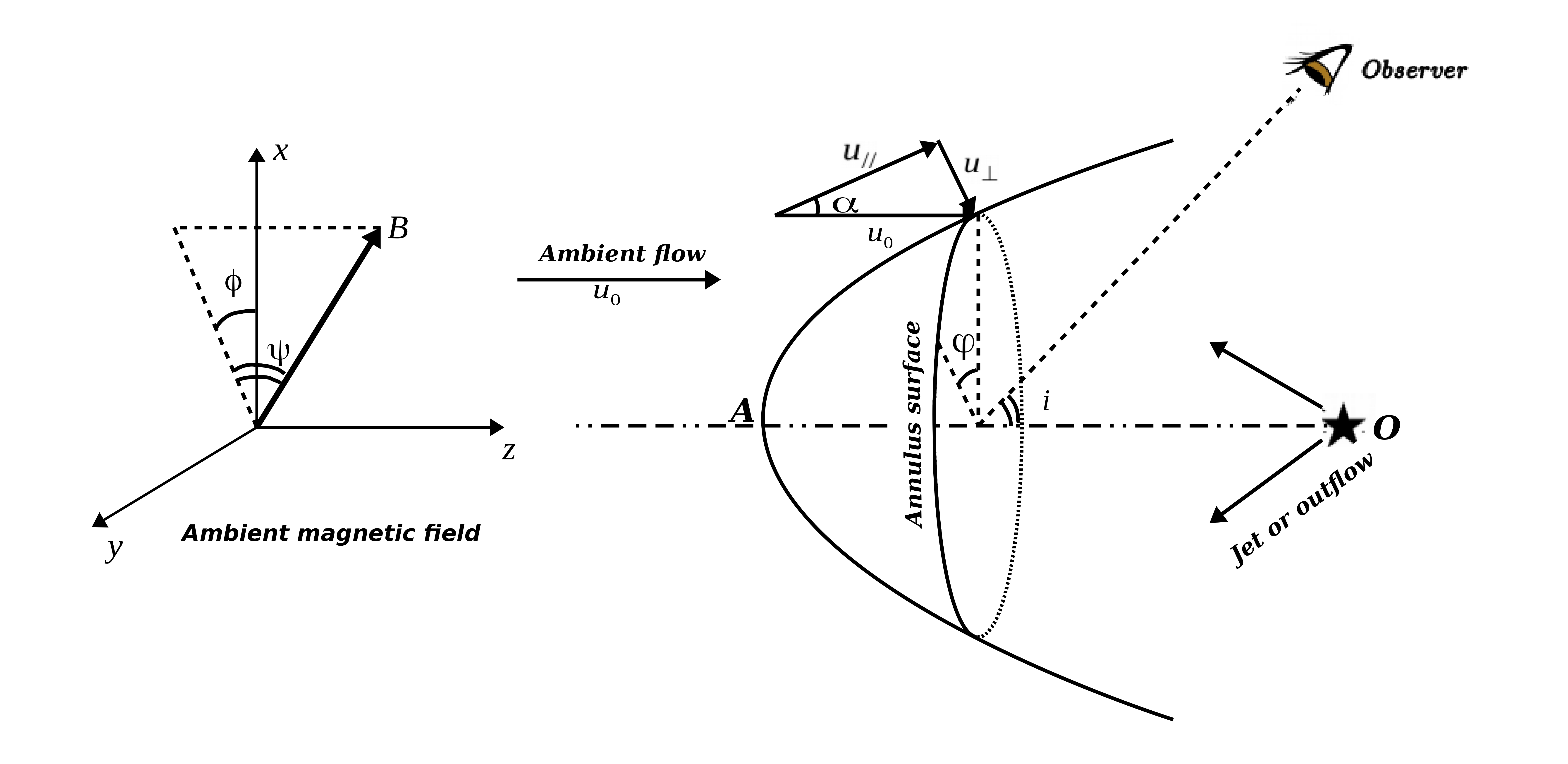}
 		\caption[shock]{Morphology of a magnetized bow shock 
 		in the frame of a star or a jet. The direction of the 
 		magnetic field is expressed by the angles $\psi$ and $\phi$. 
 		The observer lies at an angle $i$ to the z-axis in the Oxz plane.}
 		\label{fig:sketch_bow_shock}
\end{figure}

We consider an axisymmetric 3D bow shock around a supersonic star (or
a jet) travelling at the speed of $-\textbf{u}_{0}$ relative to an
ambient molecular cloud assumed to be at rest. In the frame of the
star, the impinging velocity is therefore uniform and equal to
$\textbf{u}_{0}$. The apex of the bow shock is at position A and the
star (or a reference point in the jet) at position O
(figure~\ref{fig:sketch_bow_shock}). The axis of
symmetry chosen as the z-axis is therefore along the direction (AO). The observer is assumed to lie in the (Oxz) plane and
the y-direction is chosen such that (Oxyz) is direct. The axisymmetric shape of the
bow shock is completely determined by the function $x=f(z)$.  The local position
along the planar shock can be specified by the angle between the
incoming flow and the tangent to the surface $\alpha =
\arccos(u_{\bot}/u_{0})$ \cite[see][figure 1]{S90}, and by the angle $\varphi$ between the radius and
the x-axis in the (xy) plane of projection.

The impinging velocity can be expressed  as $\textbf{u}_{0} = \hat{\textbf{t}} u_{\parallel} + \hat{\textbf{n}} u_{\bot} = u_{0}(\hat{\textbf{t}} \cos\alpha + \hat{\textbf{n}} \sin\alpha)$, where $\hat{\textbf{n}}(-\cos\alpha \cos\varphi,-\cos\alpha \sin\varphi, \sin\alpha)$ is the unit normal vector pointing inside the bow and $\hat{\textbf{t}}(\sin\alpha \cos\varphi, \sin\alpha \sin\varphi, \cos\alpha)$ is the unit tangent vector along the working surface. The effective shock speed at the local point is $v_s=u_{\bot}=u_0 \sin \alpha$. Away from the axis of symmetry, the effective entrance velocity into the shock decreases down to the sound speed $c_s$ in the ambient medium. Beyond this point, the shock working surface is a cone of opening angle $\alpha_0=\arcsin(c_s/u_0)$, wider as the terminal velocity is closer to the sound speed. In this paper, we mainly focus on the ``nose'' of the bow shock where $u_{\bot}>c_s$, and we neglect the emission from these conical ``wings'', or we simply assume that they fall outside the observing beam.

The orientation of the line-of-sight of the observer in the $(x,z)$ plane is defined by the inclination angle $i$: $\hat{\textbf{l}}(\sin i, 0, \cos i)$. The ambient uniform magnetic field is identified by the obliqueness $\psi$ and the rotation $\phi$:  $\textbf{B}/ B_0 = (\cos \psi \cos\phi,\cos\psi \sin\phi, \sin\psi)$. For each bow shock, $\psi$ and $\phi$ are fixed. 

\subsection{Distribution function of the local planar shock velocity} \label{sec:PDF}
This section aims at computing the fraction $P(u_{\bot})\mathrm{d}u_{\bot}$ of planar shocks
with an entrance planar shock $u_{\bot}$ within
$\mathrm{d}u_{\bot}$ in a given bow shock shape. This will help us
building a model for the full bow shock from a grid of planar shocks.

Considering the shock geometry as prescribed in section \ref{sec:geometry_shock}, we aim at obtaining the formula for the unit area $ds$
corresponding to these shocks as a function of $du_{\bot}$. 

The norm of a segment $dl$ on the $(x,z)$ section of the bow shock surface is: 
\begin{equation} \label{eq:dl}
	dl = \sqrt{dx^{2} + dz^{2}} = \sqrt{1 + f'^{2}(x)}dx.
\end{equation} 

Now, we take that segment and rotate it  around the $z$-axis, over a circle of radius $x$. The
area ($ds$) of the bow shock's surface swept by this segment can be expressed as:
\begin{equation} \label{eq:ds}
	ds = 2\pi x dl = 2\pi x \sqrt{1 + f'^{2}(x)}dx \mbox{.}
\end{equation}

Note that the angle $\alpha$ defined in figure~\ref{fig:sketch_bow_shock}
is also the angle between the segment $dl$ and the differential
length $dz$ along the $z$-axis. Then, the tangent of the angle $\alpha$ can 
be set as
\begin{equation} \label{eq:tan_alpha}
	\tan \alpha = \frac{dx}{dz} = \frac{1}{f'(x)} \mbox{.}
\end{equation}

In all generality, the relationship between $\alpha$ and $u_{\bot}$ will be
realized according to whether we consider the shock in the ambient medium or in the
stellar wind or jet. Then,  $ds$ as a function of $du_{\bot}$ can be obtained by
replacing that relation into equation~\ref{eq:ds}. 
However, we will only focus here on the bow shock in the ambient material. In that case, the norm of the
effective velocity (i.e., the effective normal velocity $u_{\bot}$) is 
related to the norm of the incident velocity $\textbf{u}_{0}$ through
the angle $\alpha$ as
\begin{equation} \label{eq:alpha}
	u_{\bot} = u_{0} \sin \alpha \rightarrow \alpha = \arcsin(\frac{u_{\bot}}{u_{0}}) \mbox{.}
\end{equation}

Now, $x$ can be expressed as a function of $u_{\bot}$
by substituting equation~\ref{eq:alpha} into equation~\ref{eq:tan_alpha}:
\begin{equation} \label{eq:x_vs_u}
	\tan[\arcsin(\frac{u_{\bot}}{u_{0}})] = \frac{1}{f'(x)} \rightarrow x = f'^{-1}\lbrace  \cot[\arcsin(\frac{u_{\bot}}{u_{0}})] \rbrace = g(u_{\bot}) \mbox{.}
\end{equation} 

In equation~\ref{eq:ds}, the unit are $ds$ of the shock is a function of the
coordinate $x$, while in equation~\ref{eq:x_vs_u}, the coordinate $x$ is a
function of the effective shock velocity $u_{\bot}$. To sum up, we
can obtain $ds$ as a function of $u_{\bot}$:
\begin{equation}\label{eq:ds_ambient}
	\begin{split}
	ds(u_{\bot}) & = 2\pi g(u_{\bot}) \sqrt{1 + \cot^{2}[\arcsin(\frac{u_{\bot}}{u_{0}})]}g'(u_{\bot})du_{\bot} \\ 
	            & = \pi \sqrt{1 + \cot^{2}[\arcsin(\frac{u_{\bot}}{u_{0}})]}\ d[g^{2}(u_{\bot})]
	\end{split}
\end{equation}

Finally, the distribution function of shock velocities is simply defined as
\begin{equation} \label{eq:pdf} P(u_{\bot}) =
  \frac{ds(u_{\bot})}{\int_{c_{s}}^{u_{0}}ds},
\end{equation}
so that the integral of $P(u_{\bot})$ is normalized to unity. Note that the lower limit of the integral is the sound speed in the ambient medium. This implicitly assumes that we only focus on the ``nose'' of the bow shock, where $u_\bot<c_{s}$. One could include the conical ``wings'' by adding a Dirac distribution $\delta(u_\bot=c_s)$. Conversely, one could also narrow down the integration domain if the beam intersects a smaller fraction of the bow. We implemented this mathematical formulation numerically to compute the distribution $P$ from an arbitrary input function $f$. We obtained results that agree with those obtained using the analytical expressions when the shape assumes a power-law dependence $z \sim x^\beta$.

\begin{figure}
   \begin{minipage}[c]{.48\textwidth}
      \includegraphics[width=1\linewidth, height=0.23\textheight]
      {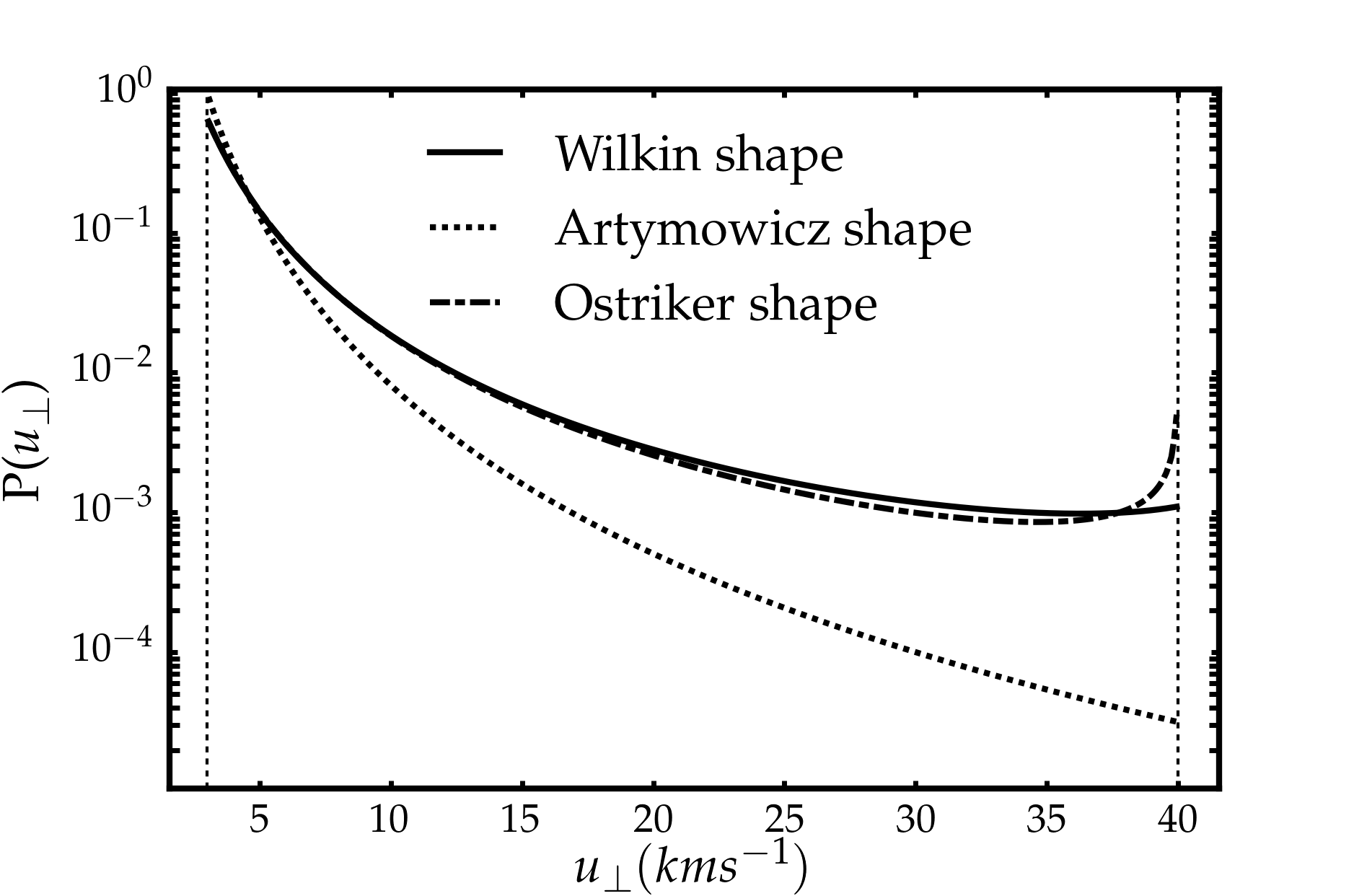}
      \caption{Statistical distributions of 1D planar shock along 
      the bow shock obtained for various bow shock shapes. 
      These distributions are dominated by low-velocity shocks.} 
	  \label{fig:pdf}
	\end{minipage}	

  \begin{minipage}[c]{.48\textwidth}
      \includegraphics[width=1\linewidth, height=0.23\textheight]
      {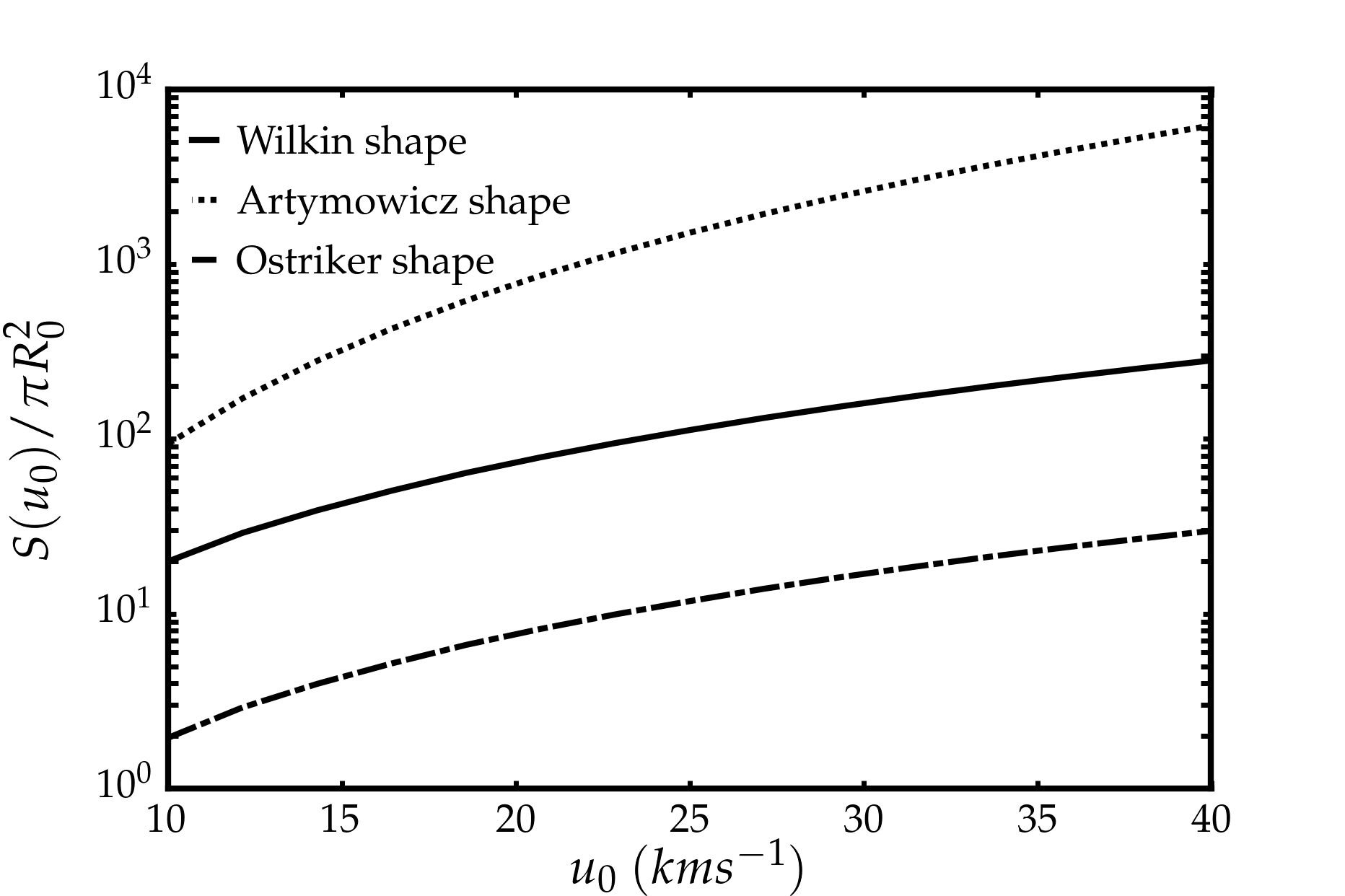}
	  \caption{Total surface of the bow shock for various bow shock 
	  shapes and terminal velocities, in units of $\pi R_0^2$ 
	  where $R_0$ is the length-scale parameter of the bow 
	  (on the order of the nose's curvature radius). } 
	  \label{fig:S}
	\end{minipage}
\end{figure}

\subsection{Example of bow shock shapes} \label{sec:shape}

In an elegant and concise article, \cite{W96} derived an analytical
description of the shape of a bow shock around a stellar wind when it is dominated by the
ram pressure of the gas. When dust grains control
the dynamics of the gas, the main forces are the gravitation pull and
the radiation pressure from the star and the shape of the shock should
then be very close to the grains avoidance parabola derived in \cite{A97}.  In fact, the ISM is a mixture between
gas and dust grains, so the actual bow shock shape should lie 
in-between.  

For the dust dominated case, the bow shock shape is
the Artymowicz parabola expressed as $z = \frac{1}{4R_{0}}x^{2} -R_{0}$ with $R_{c} = 2R_{0}$ the curvature radius at apex, $R_0$ being the star-apex distance. In the gas
dominated case, the bow shock shape follows the Wilkin formula $R =
\frac{R_{0}}{\sin \theta}\sqrt{3}\sqrt{1-\theta \cot \theta}$ with $R_{c} = 5/4R_{0}$ the curvature radius.

Finally, in the case of the tip of a jet, \cite{Ostriker01} showed that the shape of the bow shock should be cubic $z=x^{3}/R_0^2 - z_{j}$ with an infinite curvature radius (and $R_0$ and $z_j$ are length-scales parameters). Figure~\ref{fig:pdf} displays the distributions obtained for
various bow shock shapes.  Note that low-velocity shocks (u$_{\bot}$ $\leq$ 15 km$\,$s$^{-1}$) always dominate
the distribution: this stems from the fact that the corresponding surface increases further away from the
axis of symmetry, where entrance velocities decrease. The distribution for the cubic shape has a spike
due to its flatness (infinite curvature radius) near the apex. The Wilkin shape has a cubic tail but a parabolic nose. 
 In figure~\ref{fig:S} we display the dimensionless surface $S/ \pi R_0^2$ where $S$ is the total surface of the bow shock. $R_0$ is an estimate of the radius of the nose of the bow. For elongated shapes such as the parabolic shape, the total surface can be much bigger than the nose cross-section  $\pi R_0^2$. We will subsequently essentially consider an ambient shock with a parabolic shape (Artymowicz shape), unless otherwise stated.

\subsection[]{Orientation of the magnetic field} \label{sec:b}
\begin{figure*}
	\centering
	\subfloat{
      	\includegraphics[height=0.23\textheight]
      	{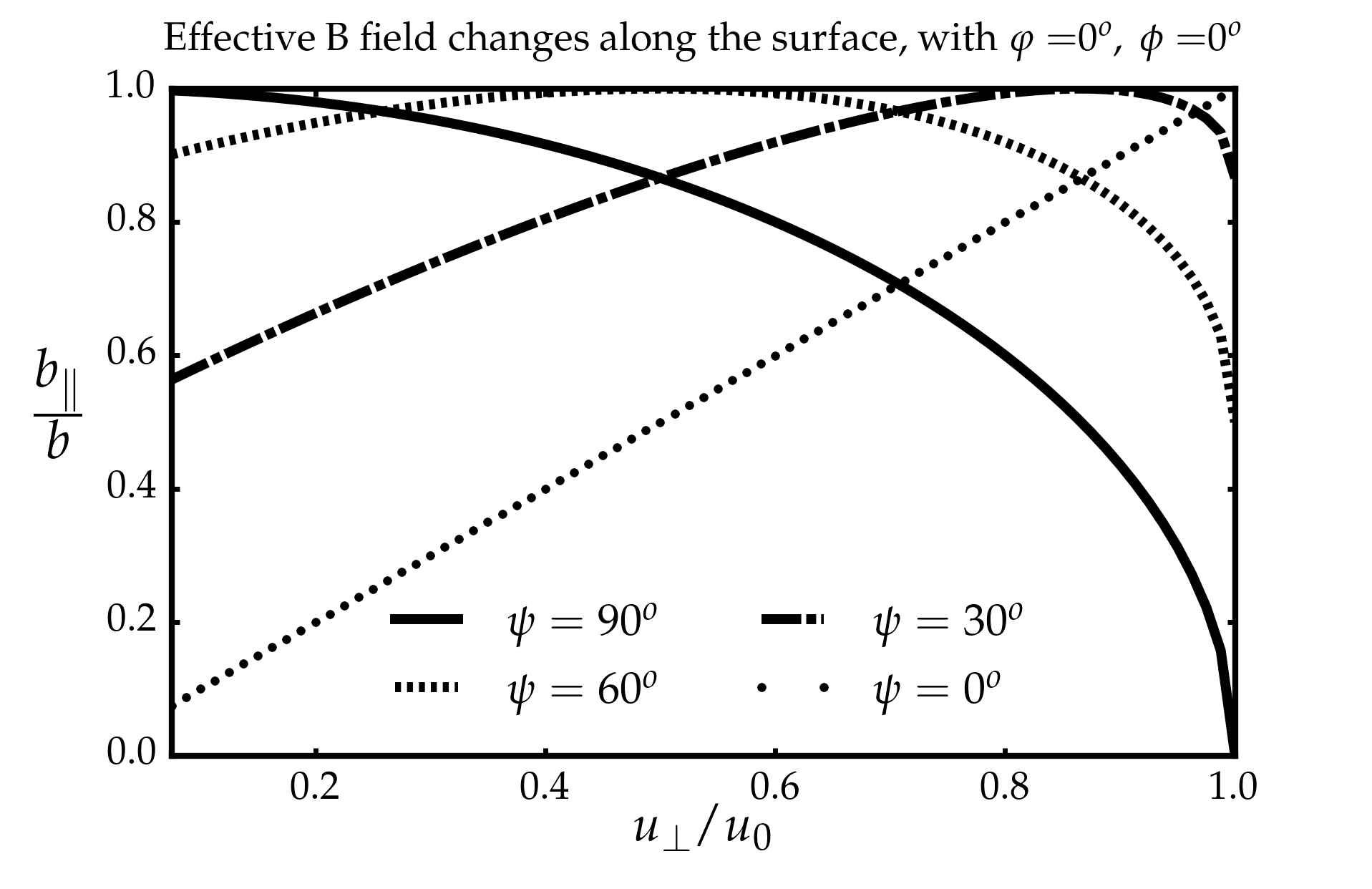}
    }
    \subfloat{
	    \includegraphics[height=0.23\textheight]
	    {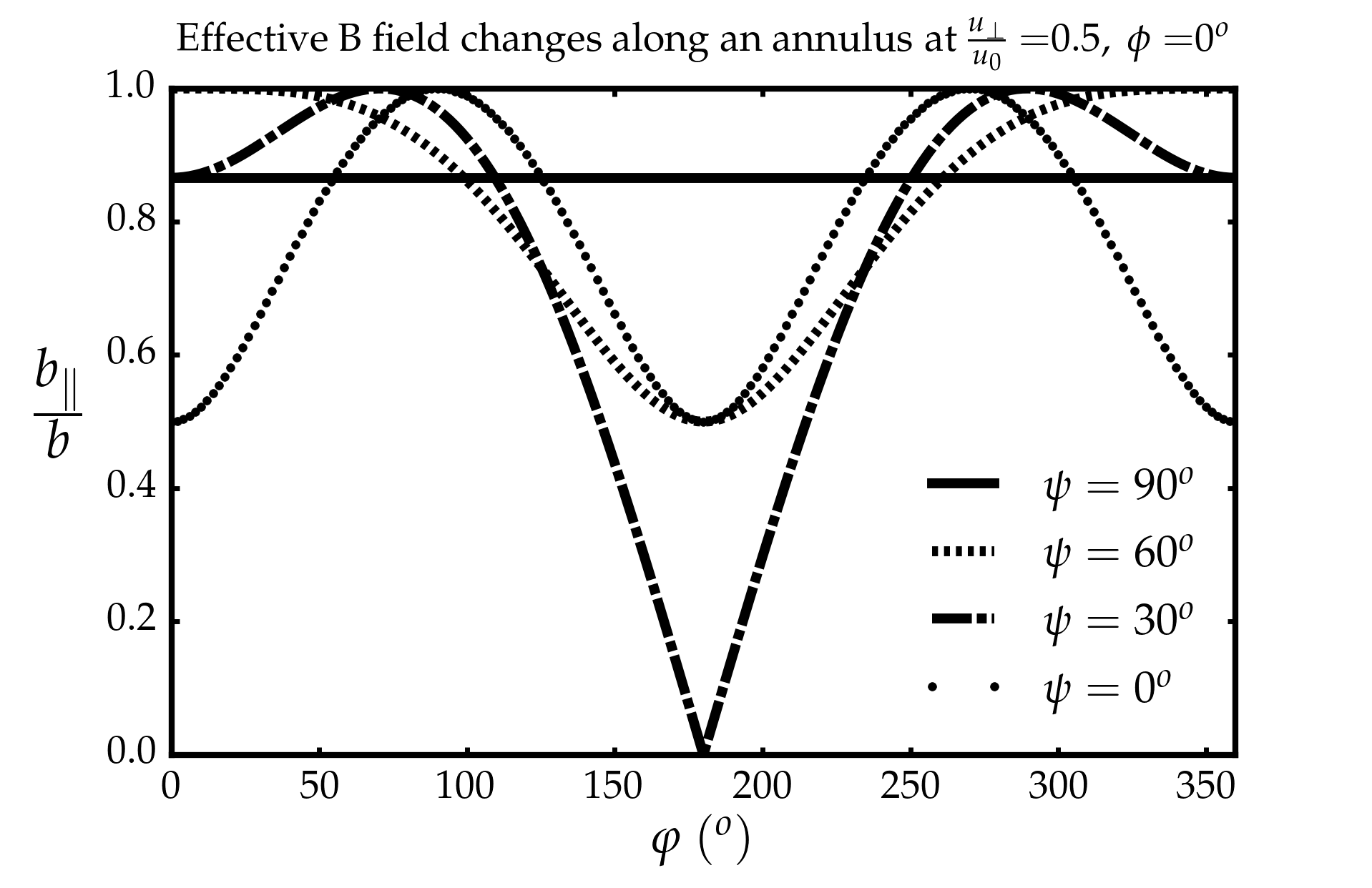}
	}
	\caption{Variation of the effective transverse magnetic field 
	$b_{\parallel}$ along the bow shock surface: for various directions 
	of \textbf{b} (left)  and based on the position of fixed direction 
	of \textbf{b} on an annulus circle (right). } 
	\label{fig:orientation_b}
\end{figure*}

The magnetic field  decouples the ions from the neutral fluid in the shock. However, as discussed in \cite{S92}, the effective magnetic field is the component of the field parallel to the shock surface. If the homogeneous pre-shock density is $n_H$, the strength scale factor of the ambient uniform magnetic field is defined as $b_0 = B_0(\mu G)/\sqrt{n_{H} [cm^{-3}]}$. The component of the field parallel to the working surface $b_{\parallel}$ is given by 
\begin{equation} \label{eq:b-parallel}
	\left(\frac{b_{\parallel}}{b_0}\right)^2 = \cos^2 \alpha \sin^2 (\varphi-\phi)+ [\sin \psi \sin \alpha+ \cos \psi \cos \alpha \cos (\varphi - \phi) ]^2
\end{equation}
where the angles $\alpha$ and $\varphi$ monitor the position in the bow shock (this expression is actually valid regardless of the bow shock shape).
Figure~\ref{fig:orientation_b} displays how this component ($b_{\parallel}$)  changes along the shock surface in a few cases. 

\section{1D planar shock models} \label{sec:1D_shock}
We now compute the chemical composition and the emission properties of
each local planar shock composing a bow shock. 
\subsection{Grid input parameters} \label{sec:grid_model}
We set all the parameters to values corresponding to typical conditions encountered in the molecular
interstellar gas in our Galaxy, as described in table 1.
We assume that the ambient gas is initially at chemical and thermal equilibrium and we compute this initial state as in \cite{PL13} by evolving the gas at constant density  during $10^{12}/n_H$ yr. Our initial elemental abundances in the gas, grain cores and ice mantles are the same in \cite{FP03}. We also include PAHs with ratio $n(PAH)/n_{H} = 10^{-6}$. The irradiation conditions are for a standard external irradiation field ($G_0=1$) but an additional  buffer of $A_{v 0}=0.1$, $N_0({\rm H}_2)=10^{20} $cm$^{-2}$ and $N_0(CO)=0$ cm$^{-2}$ is set between the source and the shock so that the gas is actually mainly molecular \citep[see][for details]{PL13}.  In our calculations, the atomic hydrogen fractions $n(H)/n_{H}$ are $7.85\ 10^{-2}$, $5.94\ 10^{-4}$ and  $5.89\ 10^{-6}$: they correspond to pre-shock gas densities of $10^{2}$, $10^{3}$ and $10^{6}\ cm^{-3}$, respectively. These initial conditions at steady state are then used as pre-shock conditions to compute the grid of planar shock models. 

 Our grid of models
has a range of shock velocities between 3 to 40 km$\,$s$^{-1}$ as in \cite{PL13}, with a velocity step of $\Delta u = 1$ km$\,$s$^{-1}$. However, we take into account the effect of the finite shock age by taking snapshots at 5 different values of age: $10^{2}$, $10^{3}$, $10^{4}$, and $10^{5}$ years for a density of $n_H=10^2~$cm$^{-3}$, and a hundred times shorter for a density of $n_H=10^4~$cm$^{-3}$. Note that the typical time to reach the steady-state in a C-type shock with $G_0=1$ is about $t_s=10^6 $yr$ / (n_H/10^2\mbox{cm}^{-3})$ \cite[with little or no magnetic field dependence, see][]{PL04a}. 

  The projected value of the magnetic field parallel to the shock $B_{\parallel}$ varies along the shock surface, so we need to sample the range of attainable values in our grid. The first constraint for a shock to exist is that its entrance velocity $u_\bot$ should be greater than the Alfv{\'e}n velocity $v_{A} =\frac{B_{\parallel}}{\sqrt{4\pi\rho}} \simeq b_{\parallel}\,1.85\,$km$\,$s$^{-1}$ where we defined the dimensionless value of the transverse magnetic field using the standard scaling $b_{\parallel}=B_{\parallel}/\mu $G$ / (n_H/{\rm cm}^{-3})^{1/2}$. The condition $u_\bot>v_{A}$ translates as $b_{\parallel}<u_\bot/1.85 $km$\,$s$^{-1}$, and we use as upper limit of our grid
$b_{\parallel}<b_{\parallel{\rm max}}=u_{\bot}$/3 km$\,$s$^{-1}$ (see figure~\ref{fig:grid_model}).

  Another important parameter is the magnetosonic speed in the charged fluid $v_{m} = \sqrt{c^{2}_{s} + B_{\parallel}^{2}/4\pi \rho_{c}}$ (where $c_{s}$ and $B^{2}/4\pi \rho_{c}$ are the speed of sound and the Alfvén speed of the {\it charged} fluid). The magnetosonic speed is the fastest signal speed in a partially ionized medium. Due to the low ionization degree in the molecular ISM, it is almost proportional to the local magnetization parameter: $v_{m} \simeq B_{\parallel}^2/4\pi\rho_c=b_{\parallel}v_{m1}$ where $v_{m1}$ is the magnetosonic speed obtained when the magnetization parameter is equal to unity. In our calculations, we find $v_{m1}= 18.5$  km$\,$s$^{-1}$ or $v_{m1}= 19.2$  km$\,$s$^{-1}$ for respective densities of $n_{H}= 10^{2}\ cm^{-3}$ or $n_{H}= 10^{4}\ cm^{-3}$. The charged fluid mass is dominated by the dust grains: the gas-to-dust ratio turns out to be $\rho/\rho_{d} = 180$ for the cores and mantle composition used in our simulations.

\subsection{J- and C-type shocks at early age} \label{sec:shock_classification}

\begin{figure}	
	\begin{minipage}[c]{.48\textwidth}
  	      \includegraphics[width=1.\linewidth, height=0.23\textheight]
     	  {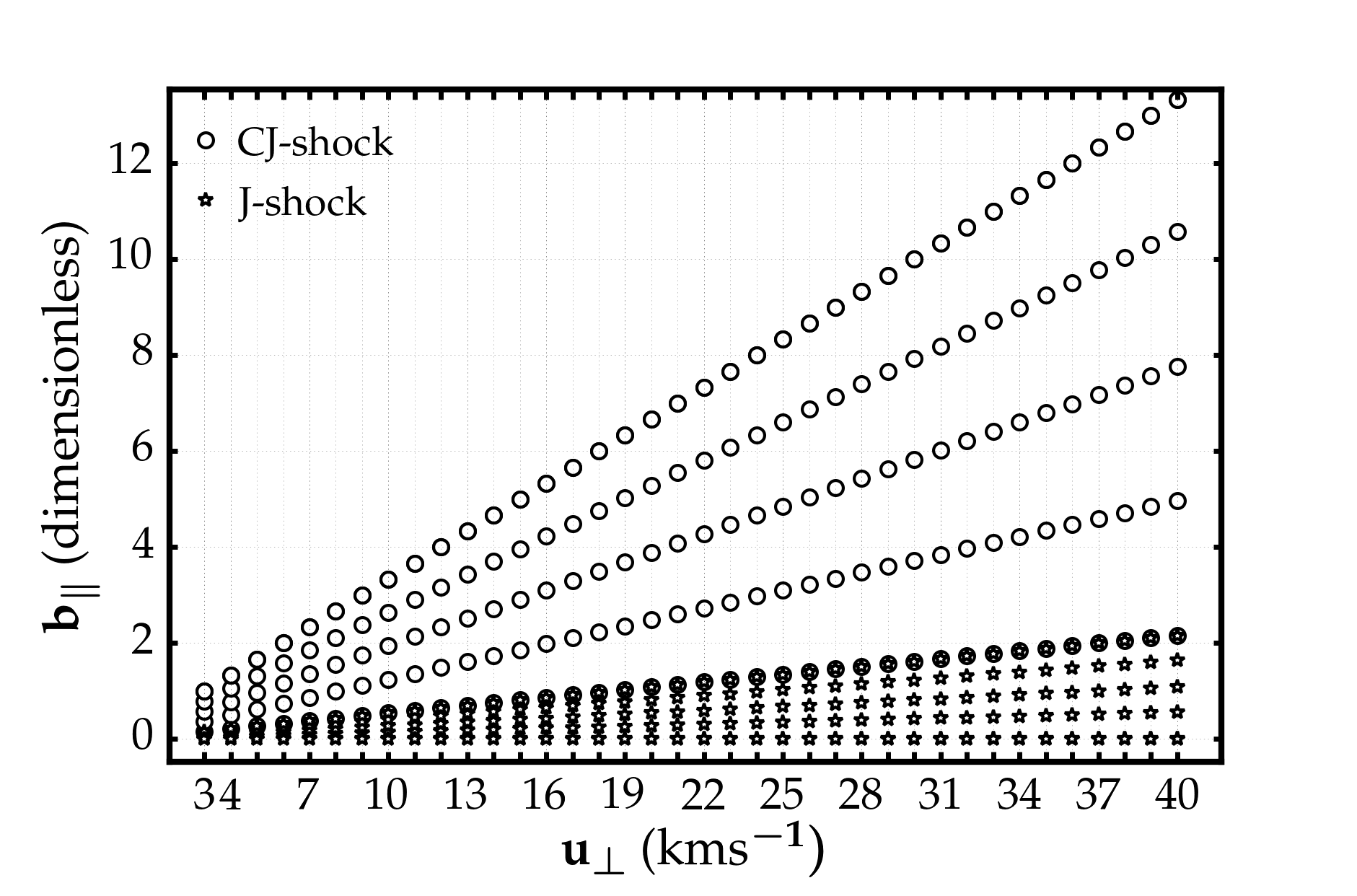}   
          \caption{Grid of 1D models in the parameter space 
                   $(u_{\bot},b_{\parallel})$.}
          \label{fig:grid_model}
     \end{minipage} 
     	
     \begin{minipage}[c]{.48\textwidth}
  	      \includegraphics[width=1.\linewidth, height=0.3\textheight]
          {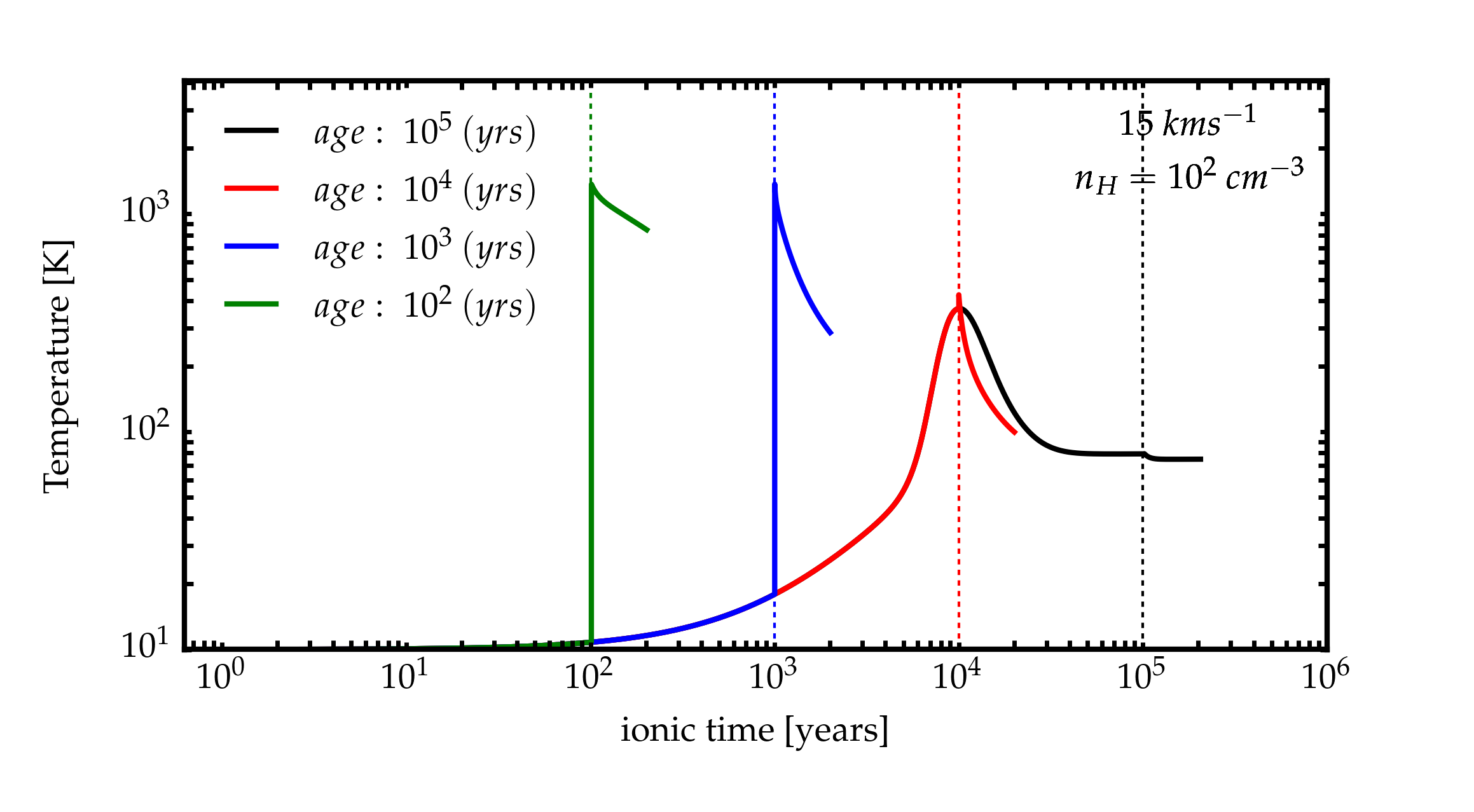}
          \caption{Temperature profile of CJ-type shocks at 
                   various shock ages for $n_H=10^{2}\ cm^{-3}$ and b=5.}
	      \label{fig:CJ-temperature}
	\end{minipage}

\end{figure}

Depending on the value  of the entrance speed relative to the entrance magnetosonic speed $v_{m}$, one can consider different kinds of shocks. When the magnetic field is weak
and/or when the ionization fraction is large, the shocks behave like hydrodynamic
shocks with an extra contribution from the magnetic pressure. Such shocks are faster than the signal speed in the pre-shock
medium. Therefore, the latter cannot "feel" the shock wave
before it arrives. Across the shock front, the variables (pressure,
density, velocity, etc.) of the fluid vary as a viscous discontinuity
jump (the so-called $J$-type shock). When the ionization fraction is small, the magnetosonic speed $v_{m}$ in the charges can be greater
than the shock entrance velocity, then a magnetic precursor forms
upstream of the discontinuity where the charged and neutral fluids dynamically decouple. The resulting friction between the two fluids
heats up and accelerates the neutral fluid. At early ages, the shock is actually composed of a magnetic precursor and a J-type tail
(it is a time-dependent CJ-type shock). \cite{Ch1998} remarked that time-dependent shocks looked like steady-state: this yielded techniques to produce time-dependent snapshots from pieces of steady-state models \citep{FP99,PL04b}. We follow the approach of \cite{PL04b} in the large compression case. The J-type front in a young C-type shock is thus inserted when the flow time in the charged fluid is equal to the age of the shock. The J-type shock ends when the total neutral flow time across the J-type part reaches the age of the shock (the same holds for young J-type shocks). As the shock gets older, the magnetic
precursor grows larger and the velocity entrance into the J-type front decreases due to the ion-neutral drag. As a result, the maximum temperature at the beginning of the J-type front decreases with age, as illustrated in figure~\ref{fig:CJ-temperature}. If the magnetic field is strong enough, the
J-type tail eventually disappears and the shock becomes
stationary. The resulting structure forms a continuous transition
between the pre-shock and the post-shock gas (a stationary C-type shock). 
  
 For each value of the entrance velocity $u_\bot$, we compute five CJ-type shock models with varying transverse magnetic field $b_{\\}$ equally spaced between 0 and $u_{\bot} / v_{m1}$, and we compute five J-type shocks models with varying transverse magnetic field $b_{\\}$ equally spaced between $u_{\bot} / v_{m1}$ and $b_{\parallel{\rm max}}=u_{\bot}$/3 km$\,$s$^{-1}$. That way, we homogeneously sample the possible shock magnetizations that are likely to occur in the 3D bow shock (see figure~\ref{fig:grid_model}). 

The main input parameters of the model are gathered 
in table \ref{tab:input_parameters}.

\begin{table*}
\centering
\begin{tabular}{l c l}
\hline \hline 
Parameter & \hspace{10mm} Value & \hspace{1mm} Note \tabularnewline
\hline 
$n_{H}$ & $10^{2}$cm$^{-3}$, $10^{4}$cm$^{-3}$  & Pre-shock density of H nuclei \tabularnewline

$A_{\nu}$ & 0.1 & Extinction shield \tabularnewline

$N_{0}(H_2)$ & $10^{20}$cm$^{-2}$ & Buffer H$_{2}$ column density \tabularnewline

$N_{0}(CO)$ & 0 cm$^{-2}$ & Buffer CO column density \tabularnewline

$G_{0}$ & 1 & External radiation field \tabularnewline

$\zeta$ & $3.10^{-17}$ $s^{-1}$ & Cosmic ray flux \tabularnewline

OPR & 3 & Pre-shock H$_{2}$ ortho/para ratio \tabularnewline

$u_{\bot}$ & $3, 4, 5, \dots 40$ km$\,$s$^{-1}$ & Effective shock velocity \tabularnewline

$b_{\parallel}v_{m1}/u_{\bot} $ & $0, \dots, 1$   & Range of $b_{\parallel}$ parameter for J-type shocks \tabularnewline

$b_{\parallel}v_{m1}/u_{\bot} $ & $1, \dots, \frac{v_{m1}}{3\mbox{km\ s}^{-1}}$ & Range of $b_{\parallel}$ parameter for CJ-type shocks \tabularnewline

age $\times n_H/$100 cm$^{-3}/$yr & $10^{2}$, $10^{3}$, $10^{4}$, $10^{5}$& Shock age \tabularnewline
\hline \hline 
\end{tabular}
\caption[]{Main input parameters of model.}
\label{tab:input_parameters}
\end{table*}

\subsection{H$_2$ excitation in C- and J-type shocks} \label{sec:CJ-excitation}

\begin{figure*}
	\centering
 	\subfloat{
 	 	\includegraphics[height=0.24\textheight]
 	 	{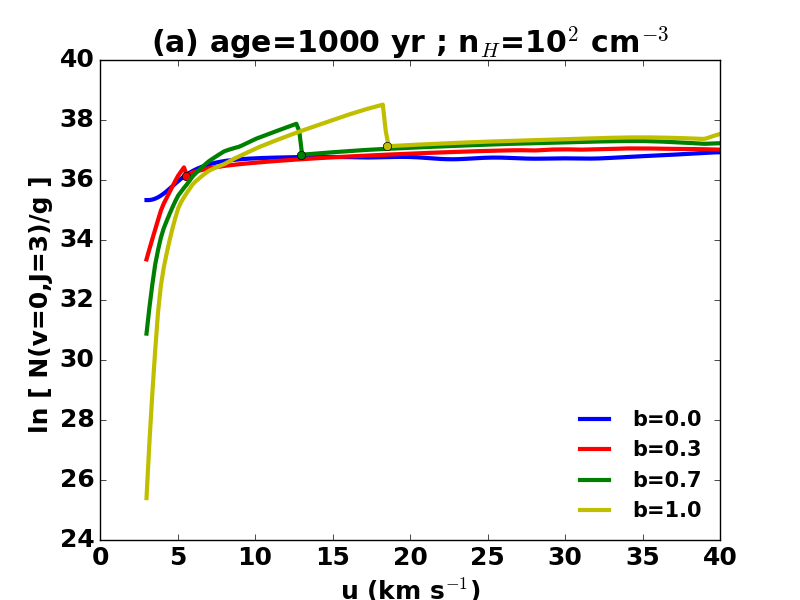}
 	}
 	\subfloat{
 	 	\includegraphics[height=0.24\textheight]
 	 	{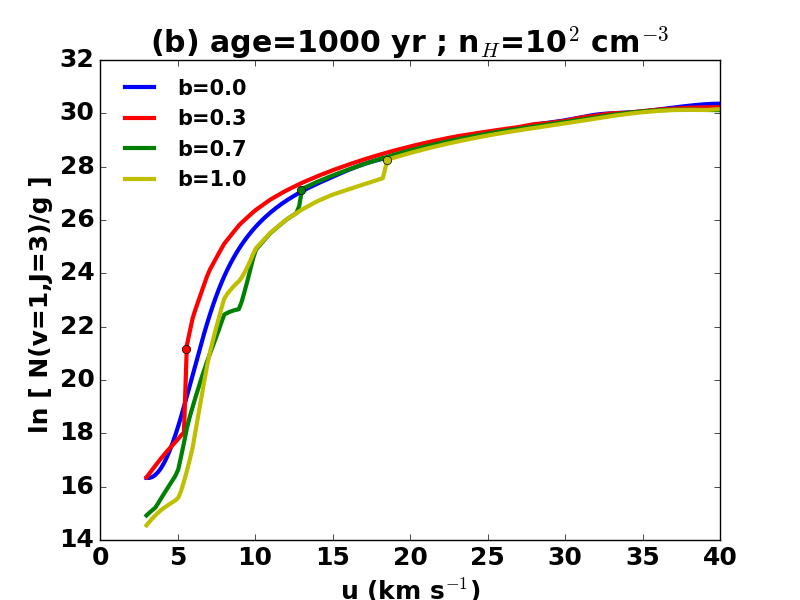}
 	} \\ 
  	\subfloat{
 	 	\includegraphics[height=0.24\textheight]
 	 	{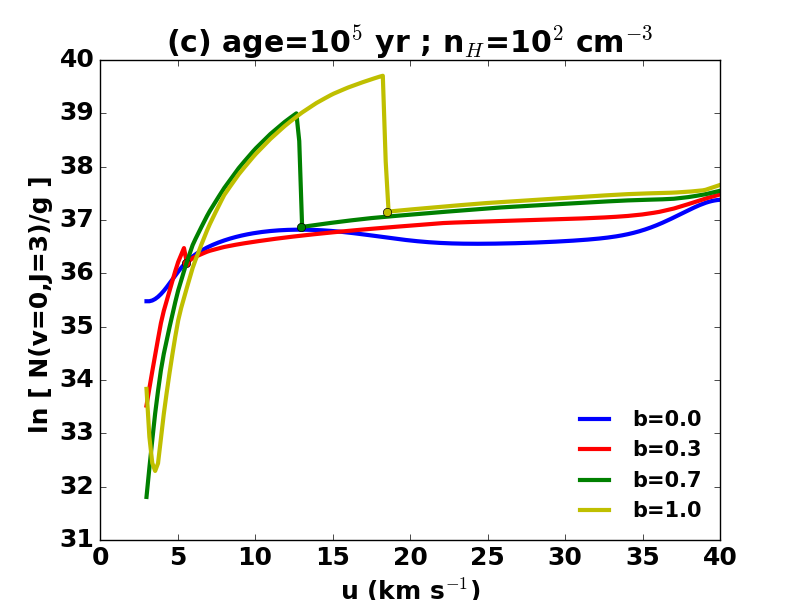}
 	}
	\subfloat{
 	 	\includegraphics[height=0.24\textheight]
 	 	{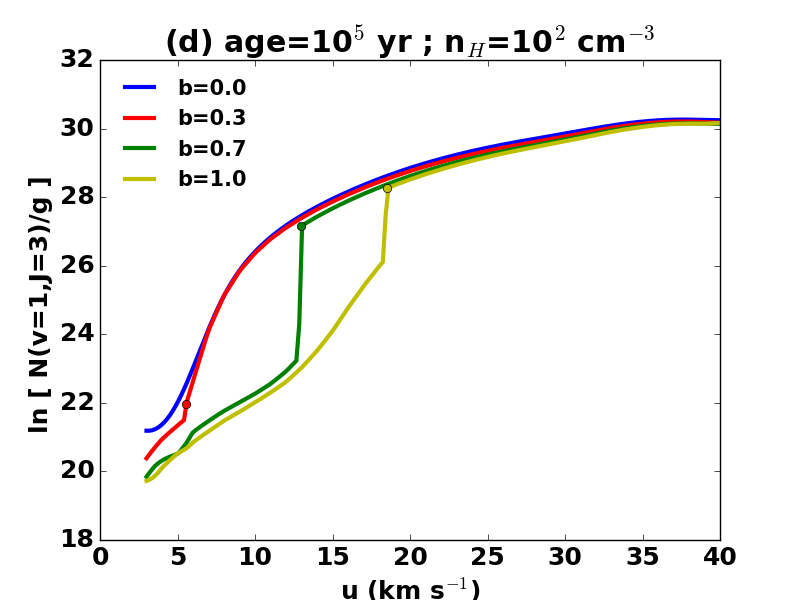}
 	} 
   \caption{Overview of results from our models for a pre-shock density 
   $n_H=10^2$/cm$^3$. We show the natural logarithm of the integrated 
   column-densities of H$_2$ populations normalized by their 
   statistical weight. They are given as a function of the velocity 
   $u$ for various values of the  magnetic field parameter $b_{\parallel}$. 
   {\it Left} panels are for the level $(v,J)=(0,3)$, the upper level of the 
   0-0S(1) line and the {\it right} panels are for the level $(v,J)=(1,3)$, 
   the upper level of the 1-0S(1) line. {\it Upper} panels are for a 
   young age of 10$^3$ yr while {\it bottom} panels are nearly 
   steady-state at an age of $10^5$ yr. In each panel, 
   the symbol 'o' marks the transition between CJ-type shocks 
   (on the left-hand side) and J-type shocks (on the right hand side), 
   when the velocity $u$ is equal to the magnetosonic speed $b v_{m1}$.}
  \label{fig:H2excited-nH1e2.png}
\end{figure*}

An H$_2$ ro-vibrational level $(v,J)$ can be populated after a collision with another species provided that the temperature yields more energy per particle than the energy level $E_{vJ}$. In a J-type shock, the sudden surge of viscous heat in the adiabatic shock front easily leads to large temperatures ($T_J=53 K (u/$km$\,$s$^{-1})^2$, see \cite{PL13}, equation 10) which are able to excite high energy levels. Figure~\ref{fig:H2excited-nH1e2.png}(a-b) show the level populations for young ages, where even CJ-type shocks are dominated by their J-type tail contribution. These figures illustrate the threshold effect for two different energy levels: their population rises quickly and reaches a plateau when $u>u_{vJ}$, with $u_{vJ}$ a critical velocity depending on the energy level. Note the weak dependence of the plateau on the shock magnetization for J-type shocks, as magnetic pressure only marginally affects their thermal properties. The critical velocity $u_{vJ}$ mainly depends on the energy level ($E_{vJ}\simeq k_BT_J$) and only weakly depends on the magnetization. 

On the other hand, C-type shocks dissipate their energy through ion-neutral friction, a process much slower than viscous dissipation: at identical velocity, C-type shocks are much cooler than J-type shocks, but their thickness is much larger. C-type shocks dominate the emission of old CJ-type shocks, when the J-type front contribution almost disappears (figure~\ref{fig:H2excited-nH1e2.png}c of the 'o' symbols). Due to their low temperature, high energy levels can never be populated (figure~\ref{fig:H2excited-nH1e2.png}d). This enhances the threshold effect, with a discontinuous jump at $u=b v_{m1}$. On the contrary, energy levels of energy lower than $k_B T_C$, with $T_C$ the typical temperature of a C-type shock, will be much more populated in a C-type shock than in a J-type shock due to the overall larger column-density. This is illustrated in the figure~\ref{fig:H2excited-nH1e2.png}c for a low energy level. The discontinuous jump at  $u=b v_{m1}$ becomes a drop instead of a surge and a peak appears in the level population. Magnetization in C-type shocks controls the compressive heating which, in turn, impacts the temperature: excitation of  H$_2$ low-energy levels in C-type shocks decreases systematically with larger magnetization, but the effect remains weak within C-type shocks. However, the magnetization is important insofar as it controls the transition between C-type and J-type shocks, which have very different emission properties.

 To summarize, at a density of $n_H=10^2$/cm$^3$, the excitation of a given H$_2$ level follows a threshold in velocity after which a plateau is reached, with little or no magnetic field dependence. However, low energy levels at old ages, for velocities below the magnetosonic speed, can be dominated by C-type shock emission. In that case, the H$_2$ level population peaks at the magnetosonic speed before reaching a plateau. Therefore, H$_2$ emission in bow shocks is likely to be mostly dominated by J-type shocks. 

\begin{figure*}
	\centering
	\subfloat{
 	 	\includegraphics[height=0.24\textheight]
 	 	{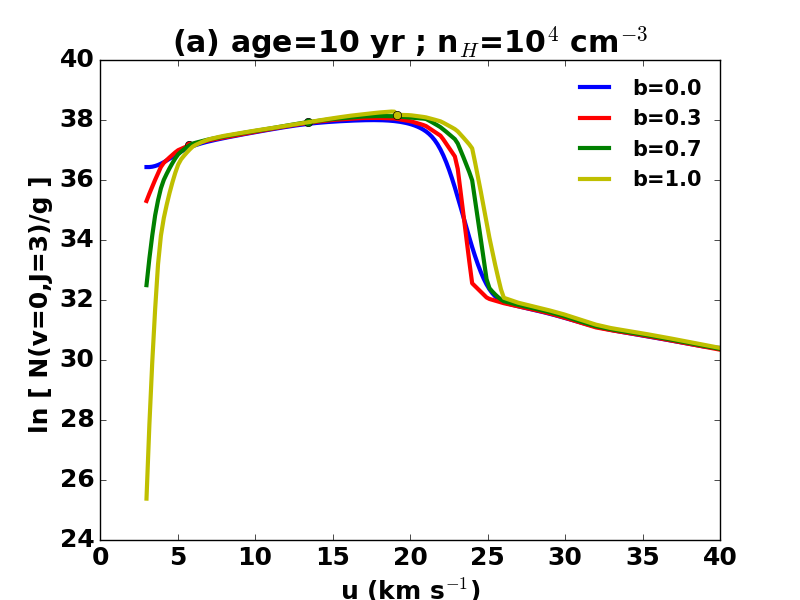}
 	}
 	\subfloat{
 	 	\includegraphics[height=0.24\textheight]
 	 	{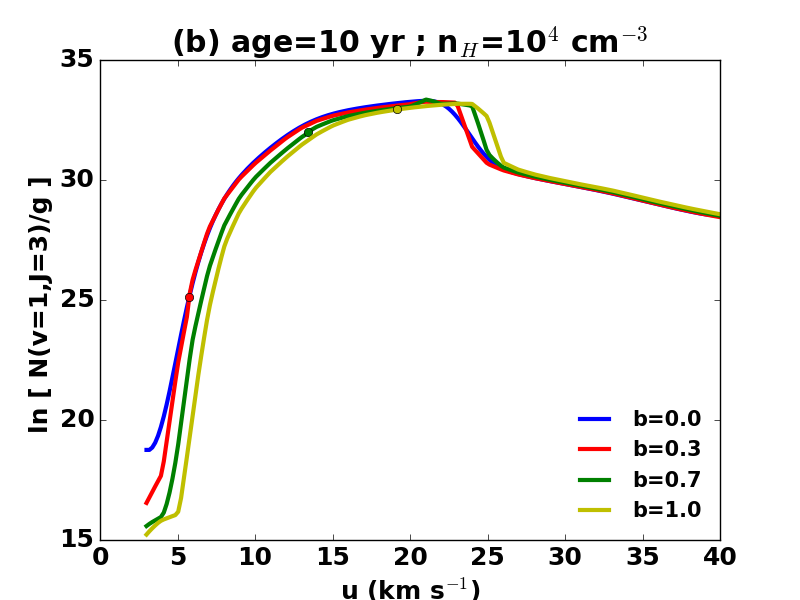}
 	} \\ 
  	\subfloat{
 	 	 \includegraphics[height=0.24\textheight]
 	 	 {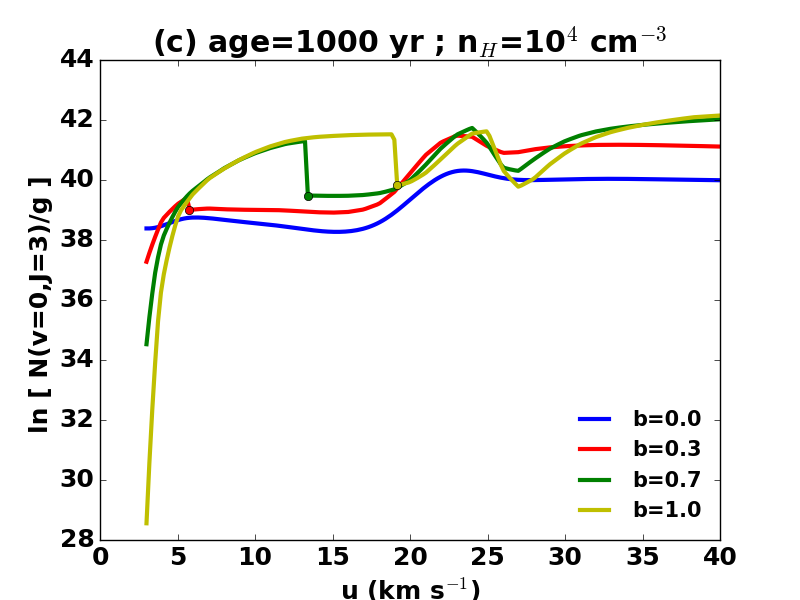}
 	}
 	\subfloat{
 	 	\includegraphics[height=0.24\textheight]
 	 	{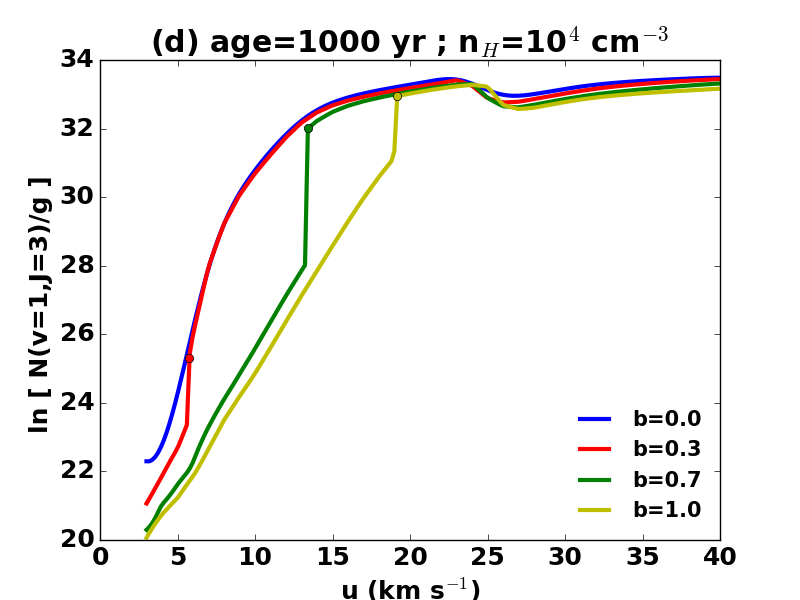}
 	}
  	\caption{Same as \ref{fig:H2excited-nH1e2.png} but for 
  	the denser case $n_H=10^4$/cm$^3$. 
  	The corresponding ages are: {\it upper} panels at 
  	a young age of 10 yr while the {\it bottom} panels 
  	are nearly steady-state at an age of $10^3$ yr.}
  \label{fig:H2excited-nH1e4.png}
\end{figure*}

  At high density, the picture is essentially unchanged, except for the effect of H$_2$ dissociation which is felt when the velocity is larger than the H$_2$ dissociation velocity: the value of the plateau decreases beyond this velocity (see the right half of each panel in figure~\ref{fig:H2excited-nH1e4.png}, which is in other respects similar to figure~\ref{fig:H2excited-nH1e2.png}).  At even higher densities, H$_2$ dissociation completely shuts off H$_2$ emission in J-type shocks, and we reach a situation where the bow shock emission is dominated by C-type shocks, as in \cite{Gustafsson10}.

\section{3D bow shock models}
\label{sec:3D_shock} 
In this section, we combine the grid of planar shocks and the statistics of planar shock velocity $u_{\bot}$ computed in the previous sections to produce observable diagnostics of 3D bow shocks. 
\subsection{H$_{2}$ excitation diagram}
\subsubsection{Excitation of a given H$_{2}$ level} \label{sec:H2_excitation_3D}

The average column-density of a given excited level of H$_2$ along the bow shock can be expressed as:
\begin{equation}
N^{\rm tot}_{vJ}(age,u_0,b_0,\psi)= \int_0^{2\pi}\frac{d\varphi}{2 \pi} \int_{c_{s}}^{u_{0}}  P_{u_0}(u_\bot)  N_{vJ}(age,u_{\bot},b_{\parallel}) du_{\bot}
\end{equation}
where $P_{u_0(u_\bot)}$ is the distribution computed in section 2 and $N^{\rm tot}_{vJ}$ and $N_{vJ}$ are the column-densities
of H$_{2}$ in the excited level $(v,J)$ in the whole bow shock and in each planar shock, respectively.

\begin{figure}
   \begin{minipage}[c]{.48\textwidth}
        \includegraphics[width=1\linewidth, height=0.25\textheight]
        {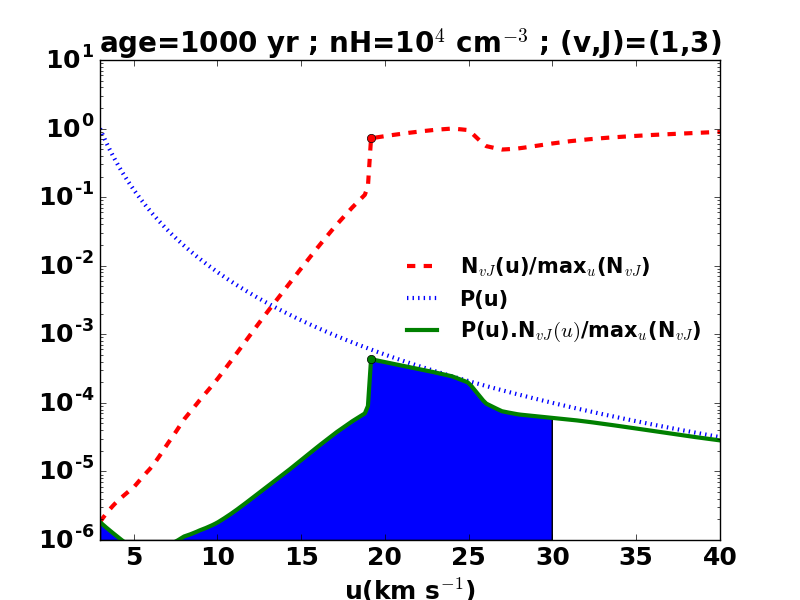}
        \caption{Illustration of the ``Gamow-peak'' effect on 
        the integration of the total column densities of the 
        H$_2$ level $(v,J)=(1,3)$ in a bow shock with 
        terminal velocity $u_0=30$ km$\,$s$^{-1}$, $n_H=10^2$/cm$^3$, 
        and the age is 10$^5$ yr.} 
        \label{fig:illustration-convolution}
   \end{minipage}     		
\end{figure}

  As noted in section \ref{sec:CJ-excitation}, $N_{vJ}$ sharply increases as a function of $u_{\bot}$ at a given threshold velocity $u_{vJ}$ before reaching a plateau. We also showed that the statistical distribution was steeply decreasing as a function of $u_{\bot}$. As a result, the product of the two peaks at around $u_{vJ}$ and its integral over $u_{\bot}$ is a step function around $u_{vJ}$ (see figure~\ref{fig:illustration-convolution}). This situation is reminiscent of the Gamow peak for nuclear reactions. Then, $N^{\rm tot}_{vJ}(u_0)$ tends to a finite value when $u_0$ is much greater than the threshold velocity $u_{vJ}$. The final value depends both on magnetization and age.

\subsubsection{Resulting H$_{2}$ excitation diagram}
The excitation diagram displays the column densities in each excited level (normalized by their statistical weight) as a function of their corresponding excitation energy. This is an observational diagnostic widely used to estimate the physical conditions in the emitting gas.

\begin{figure*}
	\centering
	\subfloat{
 	 	\includegraphics[height=0.24\textheight]
 	 	{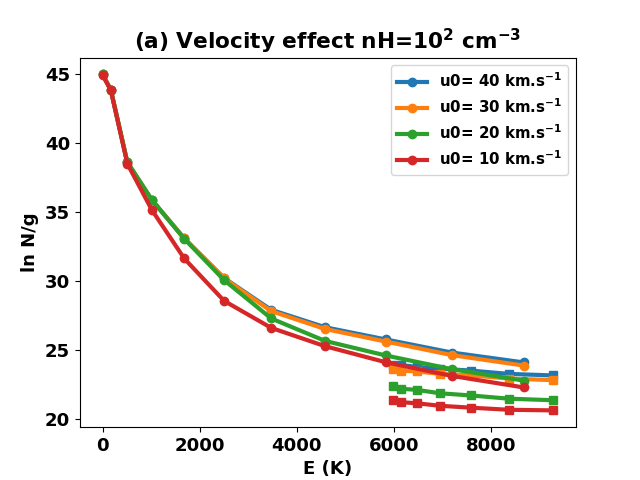}
 	}	
	\subfloat{
 	 	 \includegraphics[height=0.24\textheight]
 	 	 {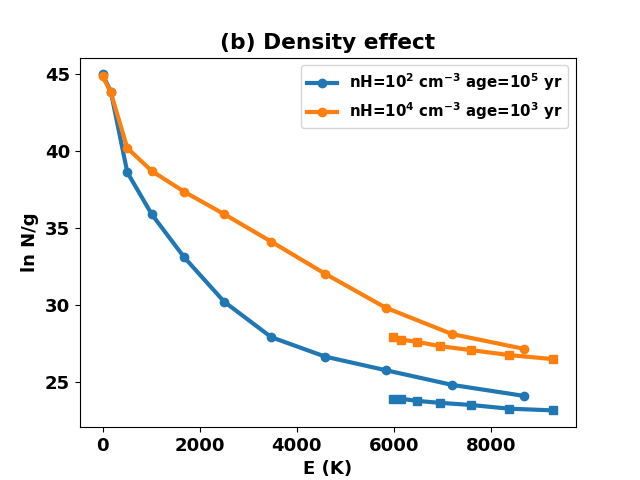}
 	} \\ 
 	\subfloat{
 	 	 \includegraphics[height=0.24\textheight]
 	 	 {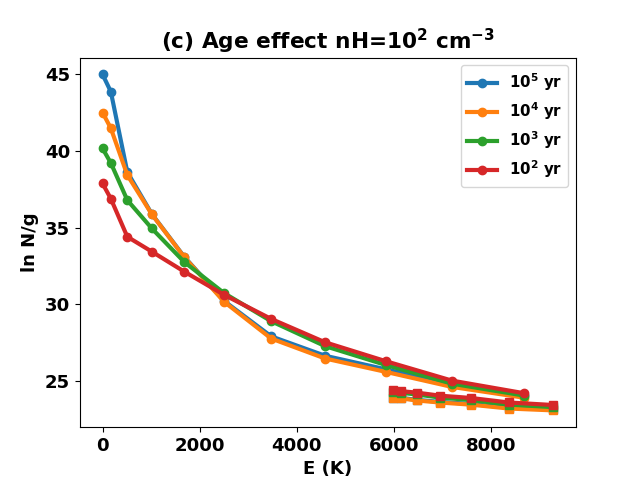}
 	} 	
 	\subfloat{
 	 	 \includegraphics[height=0.24\textheight]
 	 	 {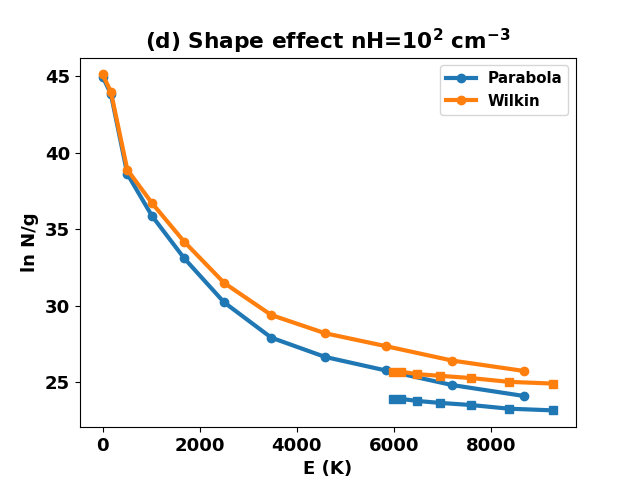}
 	} \\ 
 	\subfloat{
 	 	 \includegraphics[height=0.24\textheight]
 	 	 {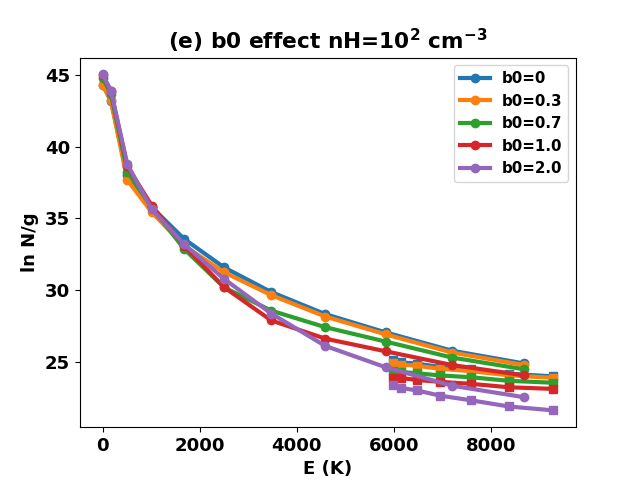}
 	}
 	\subfloat{
 	 	 \includegraphics[height=0.24\textheight]
 	 	 {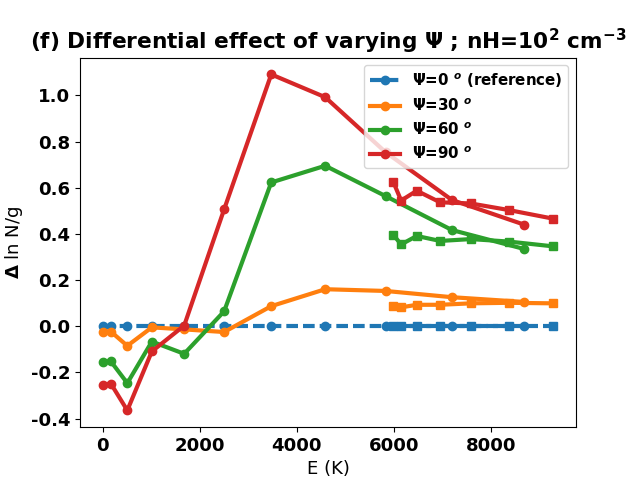}
 	}
    \caption[H2_excitation]{Excitation diagrams of H$_{2}$ 
    showing the effect of varying some of the parameters of the model. 
    The reference model is nH=1e2/cm3, age=1e5 yr, b0=1, Psi=0, 
    shape=Parabola, u0=40 km$\,$s$^{-1}$. It is always displayed in solid blue. 
    Connected circle symbols have all $v=0$ (pure rotational levels) 
    while square symbols have $v=1$.}
      \label{fig:h2_diagram}
\end{figure*}

Figure~\ref{fig:h2_diagram}a shows the influence of the terminal velocity on the excitation diagrams of H$_{2}$ at an age of $10^{4}$ yr. As expected, the excitation diagram saturates at large velocity, when $u_0$ is larger than all the individual $u_{vJ}$ of the levels considered. That saturation occurs quicker at low energy levels, as the corresponding critical velocity is lower.

Figure~\ref{fig:h2_diagram}b illustrates the effect of density on the excitation diagram.  Roughly speaking, the column-densities are proportional to the density, but in this example (40 km$\,$s$^{-1}$ bow shock), higher energy levels are subject to H$_2$ collisional dissociation, and they are slightly less populated relative to their low energy counter part.
 
At young ages, shocks are dominated by the emission properties of J-shock: as time passes, C-type shocks increase the emission of low energy levels and the excitation diagram of the bow shock is slightly steeper at the origin 
(figure~\ref{fig:h2_diagram}c).  Interestingly, the energy level just above 2000K does not seem to be affected by age (it is also weakly affected by all the other parameters, the safe density) and all the curves converge on this point.

As mentioned in section \ref{sec:shape}, the shape of bow shocks affects the velocity distribution and the relative weight between of the large velocities increases when one moves from a parabola to a Wilkin shape. As a result, a bow shock with a Wilkin shape has more excited high energy levels than a parabolic bow shock (figure~\ref{fig:h2_diagram}d).

Finally, the magnetic field tends to shift the transition between C-type and J-type shocks in the bow shock to larger velocities. At early age, it does not matter much, since both C-type and J-type shocks are dominated by J-type shock emission. At later ages, though, the low energy levels get an increasing contribution from C-type shocks and see their excitation increase.  
Conversely, high energy levels are less excited because the overall temperature of the shock decreases, as seen on figure~\ref{fig:h2_diagram}e. The orientation of the magnetic field azimuthally affects the range of values of $b$ (as $\varphi$ varies) but its main systematic effect is to shift the magnetization from low velocities to large velocities as it gets more and more parallel to the axis of symmetry (figure~\ref{fig:orientation_b}). Figure~\ref{fig:h2_diagram}f shows the differential effect caused by varying the angle $\Psi$: tending $\Psi$ to 0$^{o}$ amounts to increasing $b$ (high energy levels are less excited, whereas low energy levels are more excited). The resulting change is subtle but we show below that it might still be probed by observations.

\subsubsection{Using 1D models to fit 3D excitation diagrams}
\begin{figure*}
	\centering
	\subfloat{
		\includegraphics[height=0.22\textheight]
		{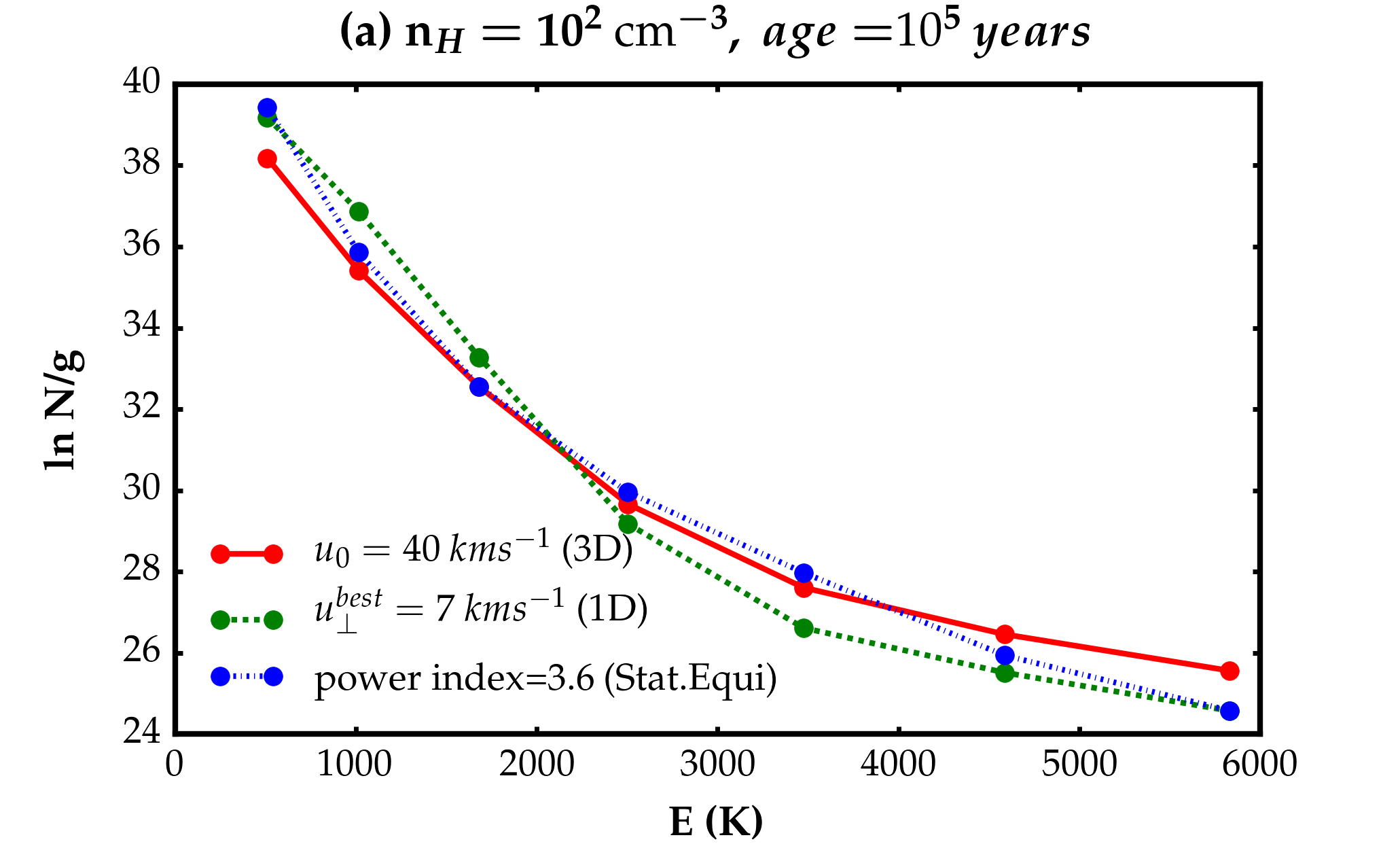}
 	}
 	\subfloat{
 		\includegraphics[height=0.22\textheight]
 		{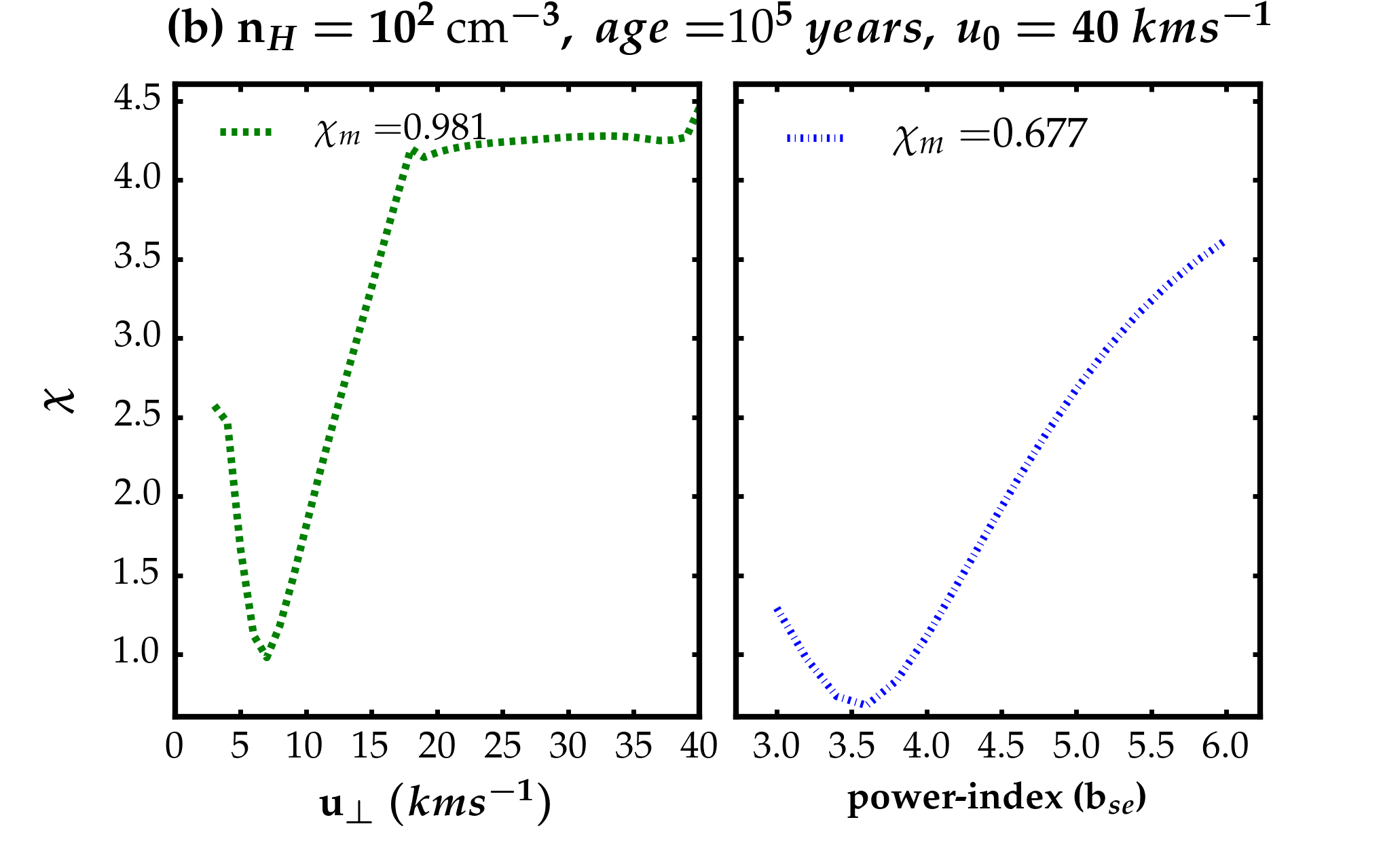}
 	}	\\ 
	\subfloat{
		\includegraphics[height=0.22\textheight]
		{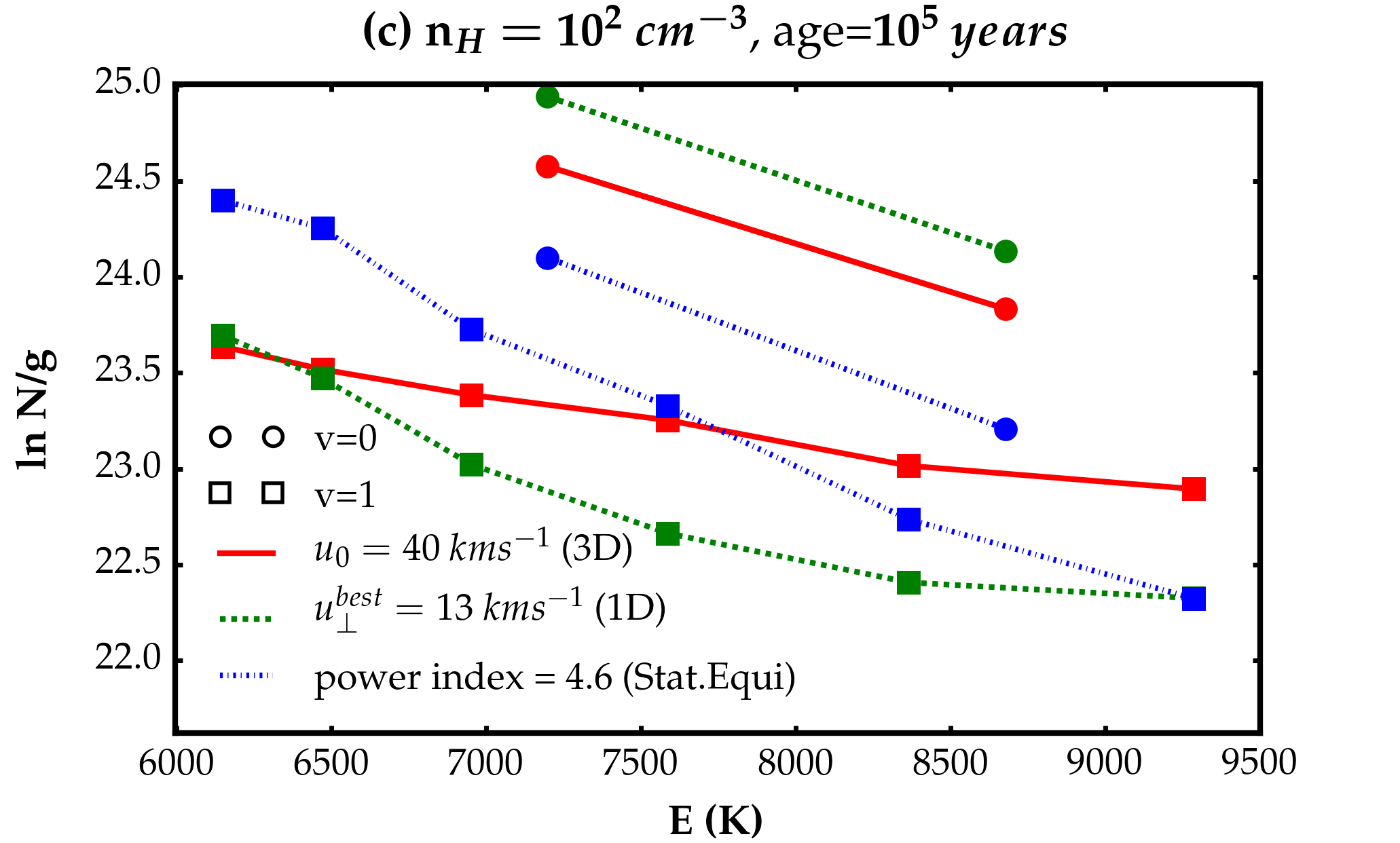}
 	}
	\subfloat{
		\includegraphics[height=0.22\textheight]
		{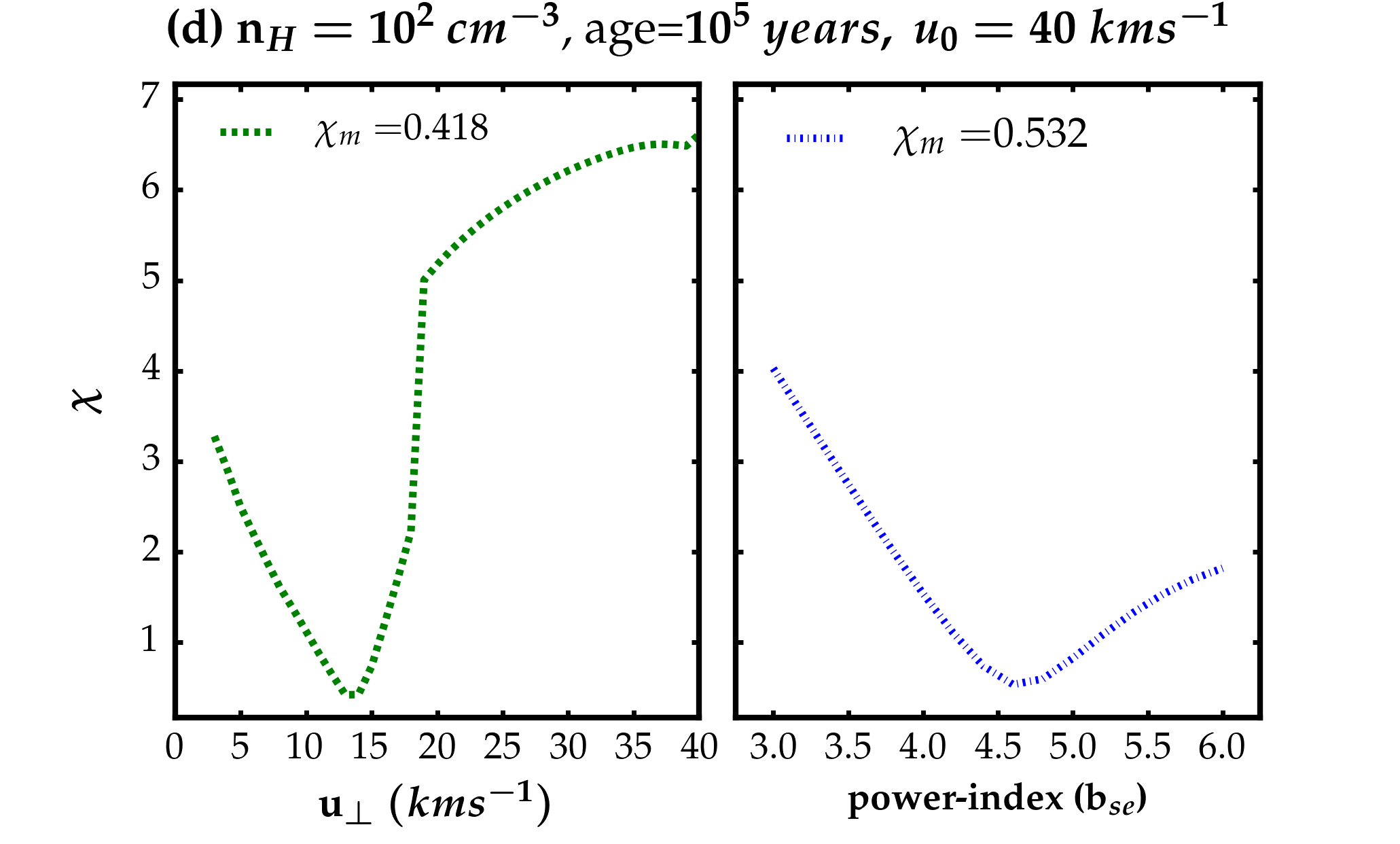}
 	}
 	\caption{Results of the fit of 1D models 
 	and statistical equilibrium approximations to a 3D bow shock. 
 	(a-b) Transitions with upper level $E_{vj} < 5900~$K (JWST-like) 
 	are used. (c-d) Fitted transitions have 5900 K$<E_{vj} <10\,000~$K. 
 	(a-c) Comparison of the 
 	excitation diagrams of the bow shock to the best 1D fit 
 	and the best NY08 fit, 
 	(b-d) standard deviation of the natural logarithm difference 
 	between the two diagrams ($\chi=\sqrt{\chi^2}$) as the entrance 
 	velocity in the 1D model 
 	and the power-index in NY08 assumption vary. 
 	The bow shock parameters are: 
 	pre-shock density $10^{2}\ cm^{-3}$, $b_0=1$, $\psi=90^o$, 
 	and the age is 10$^5$ yr. 
 	Connected circle symbols all have $v=0$ (pure rotational levels) 
    while square symbols have $v=1$.}	
      \label{fig:fit}
\end{figure*}

Observations often consider low energy transitions (pure rotational or low vib-rotational levels): although we included the first 150 levels in our calculations, here we mainly consider the levels with an energy up to $10^{4}$K. The two lowest rotational states (J=0 and 1) are, of course, unobservable in emission. The \textit{James Webb Space Telescope} (JWST) will observe pure rotational transitions up to energies of about 5900K (seven levels involved). This is similar to the performances of its predecessors: the Infrared Space Observatory (ISO) and the \textit{Spitzer} telescope. These two infrared telescopes have been used to observe shocked regions, generate excitation diagrams and maps around Young Stellar Objects (YSOs) (e.g., \citealt{Giannini04, Neufeld09}) or supernova remnants (SNRs) (e.g., \citealt{Cesarsky99, Neufeld14}) shocks. The AKARI mission has also been used for similar purposes in SNRs environments (e.g., \citealt{Shinn11}). In addition, note that the JWST will also target rovibrational transitions. Finally, the \textit{Echelon-Cross-Echelle Spectrograph} (EXES, operating between 4.5 and 28.3 microns, \citealt{Dewitt14}) onboard the \textit{Stratospheric Observatory For Infrared Astronomy} (SOFIA) should allow observations of pure rotational transitions of H$_2$, but no program has been explicitely dedicated to the observation of shocked H$_2$ with this instrument so far.

Most observations are unable to resolve all details of a bow shock, and the beam of the telescope often encompass large portions of it, therefore mixing together planar shocks with a large range of parameters. 
However, it is customary to use 1D models to interpret observed excitation diagrams. Previous work (\citealt{Neufeld08}, hereafter NY08; \citealt{Neufeld09}; \citealt{Neufeld14})  have also shown that statistical equilibrium for a power-law temperature distribution T$^{-b_{SE}}$ dT can be quite efficient at reproducing the observed H$_{2}$ pure rotational lines. We thus seek to explore how accurately these two simple models perform as compared to 3D bow shocks. We consider the worst case scenario where the whole nose of a parabolic bow shock is seen by the telescope: the effective entrance velocity $u_{\bot}$ varies from the speed of sound $c_{s}$ (in the wings of the bow shock) to the terminal velocity $u_{0}$ (at the apex of the bow shock).

The following $\chi$ function is used to estimate the distance between 1D and 3D models:

\begin{equation}
	\chi^{2} = \frac{1}{L}\sum_{vj} [\ln(\frac{N^{\rm tot}_{vj}}{g_{vj}}) - \ln(\frac{N^{u_{\bot}}_{vj}}{g_{vj}}) - C]^{2}
\end{equation}
with $L$ the number of observed vib$\_$rotational levels $(v,j)$, and $g_{vj}$ the statistical weight of each level $(v,j)$. The constant $C$ reflects the fact that the beam surface at the distance of the object may not match the actual emitting surface of the bow-shock, partly because of a beam filling factor effect and partly because the bow-shock surface is curved. We assume here that the observer has a perfect knowledge of the geometry and we take $C=0$, which means that the 1D shock model has the same surface as the 3D bow-shock to which it is compared with. The best 1D model and power-law assumption selected is the one yielding the smallest $\chi^{2}$ value on our grid of 1D models. 

  Figure~\ref{fig:fit} shows the result of the fit on a 30 km$\,$s$^{-1}$ bow shock at age $10^{5}$ years, density $n_H=10^2$ cm$^{-3}$, and magnetization parameter $b_0=1$ ($\Psi=90^o$). 1D models have the same parameters (same age, pre-shock density and $b_{\parallel}=1$) except the entrance velocity $u_\bot$. We find that the best velocity is either 8 or 13 km$\,$s$^{-1}$ depending on the range of lines considered. This is way below the terminal velocity and this illustrates again the fact that the resulting 3D excitation diagram is dominated by low velocity shocks. As a consequence, the use of higher energy lines reduces the bias, and a cubic shape for the bow shock yields less bias towards low velocity than a parabolic shape (not shown here). In the left hand sides of the panels (b)-(d), the resulting $\chi^2$ is around one in all cases: it corresponds to an average mismatch of about a factor of 3 between the 3D and 1D column-densities, a common result when comparing 1D models and observations. 

\begin{figure}
	\begin{minipage}[c]{.48\textwidth}
    	 \includegraphics[height=0.235\textheight]
    	 {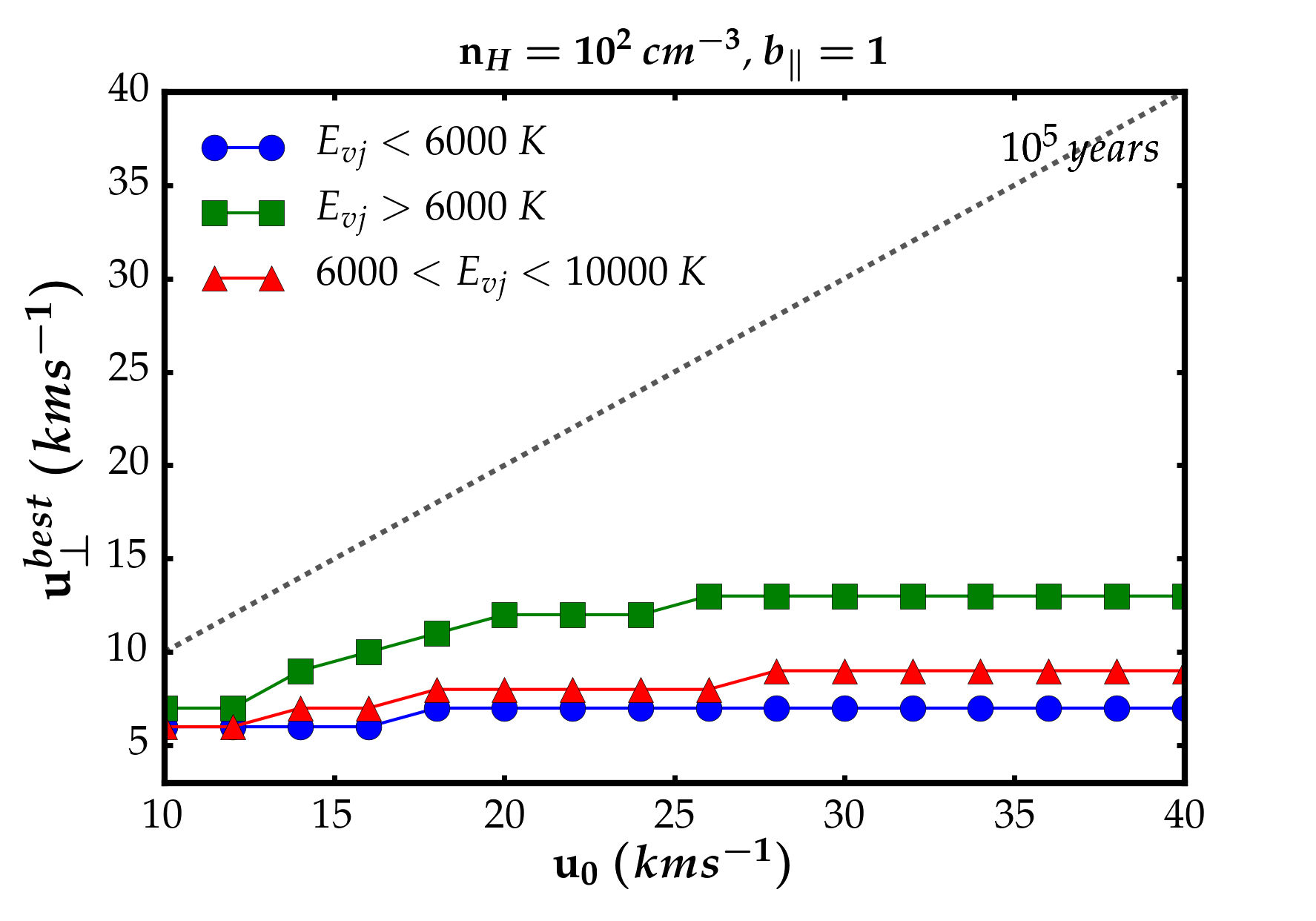}    	
     \caption{Velocity bias between 1D and 3D model. 
     Blue circle symbols fit only $E_{vj} < 5900 K (JWST-like)$, 
     the Green square symbols fit only $E_{vj} > 5900 K$ (ground based) 
     and the red triangles fit $6000 < E_{vj} < 10000K$. 
     The parameters of the bow shock 
     are the same as for figure~\ref{fig:fit}. 
     The dotted black line is $u^{best}_{\perp} = u_{0}$.}	
      \label{fig:bias_v}
  	\end{minipage}
\end{figure}

 Figure~\ref{fig:bias_v} systematically explores this bias as a function of the bow shock terminal velocity: the best 1D model usually has an entrance velocity smaller than the terminal velocity of the 3D bow-shock. Moreover, when the 3D excitation diagram saturates at large $u_{0}$, the best 1D model does not change. 
  
 Following the approach of NY08, we calculate the H$_{2}$ levels population in statistical equilibrum for a range of temperatures (100K to 4000K) and convolve this with a power-law distribution of the gas temperature. We explore power-indices ($b_{SE}$) varying from 3 to 6 (as in NY08) in steps of 0.2. We recover the fact that the NY08 approximation performs very well in the low energy regime of pure rotation. In the case displayed in figure \ref{fig:fit}(a), our best fit power-index is 3.6, close to the estimation of 3.78 for parabolic bow-shocks calculated by equation (4) in NY08. However, figure~\ref{fig:fit}(c) shows that this simple approach fails for vibrational levels, or rotational levels of higher energy.
  
\begin{figure}
   	\begin{minipage}[c]{.48\textwidth}
     	 \includegraphics[height=0.235\textheight]
     	 {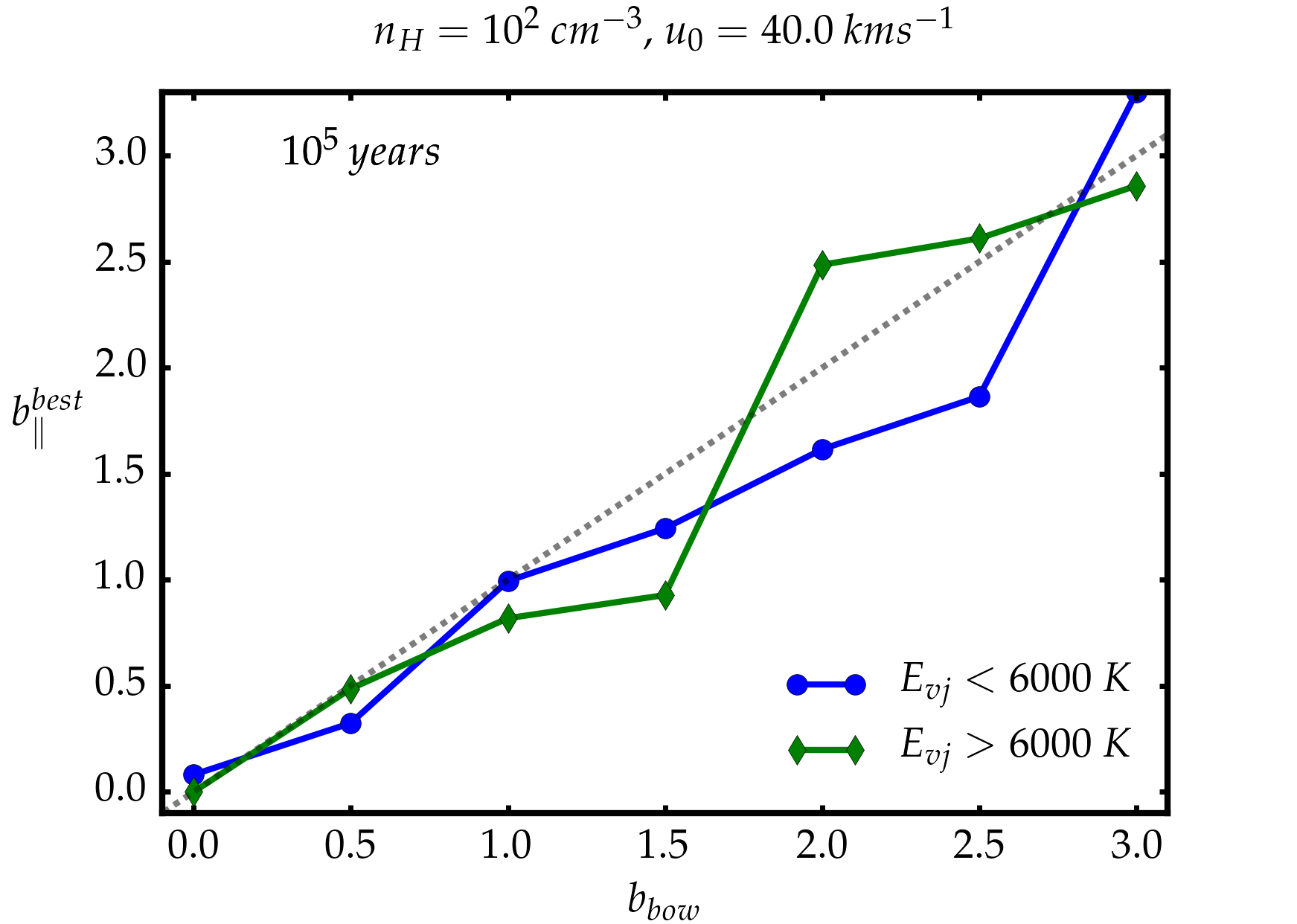}
	\end{minipage}
	\begin{minipage}[c]{.08\textwidth}
     	 \includegraphics[height=0.235\textheight]
     	 {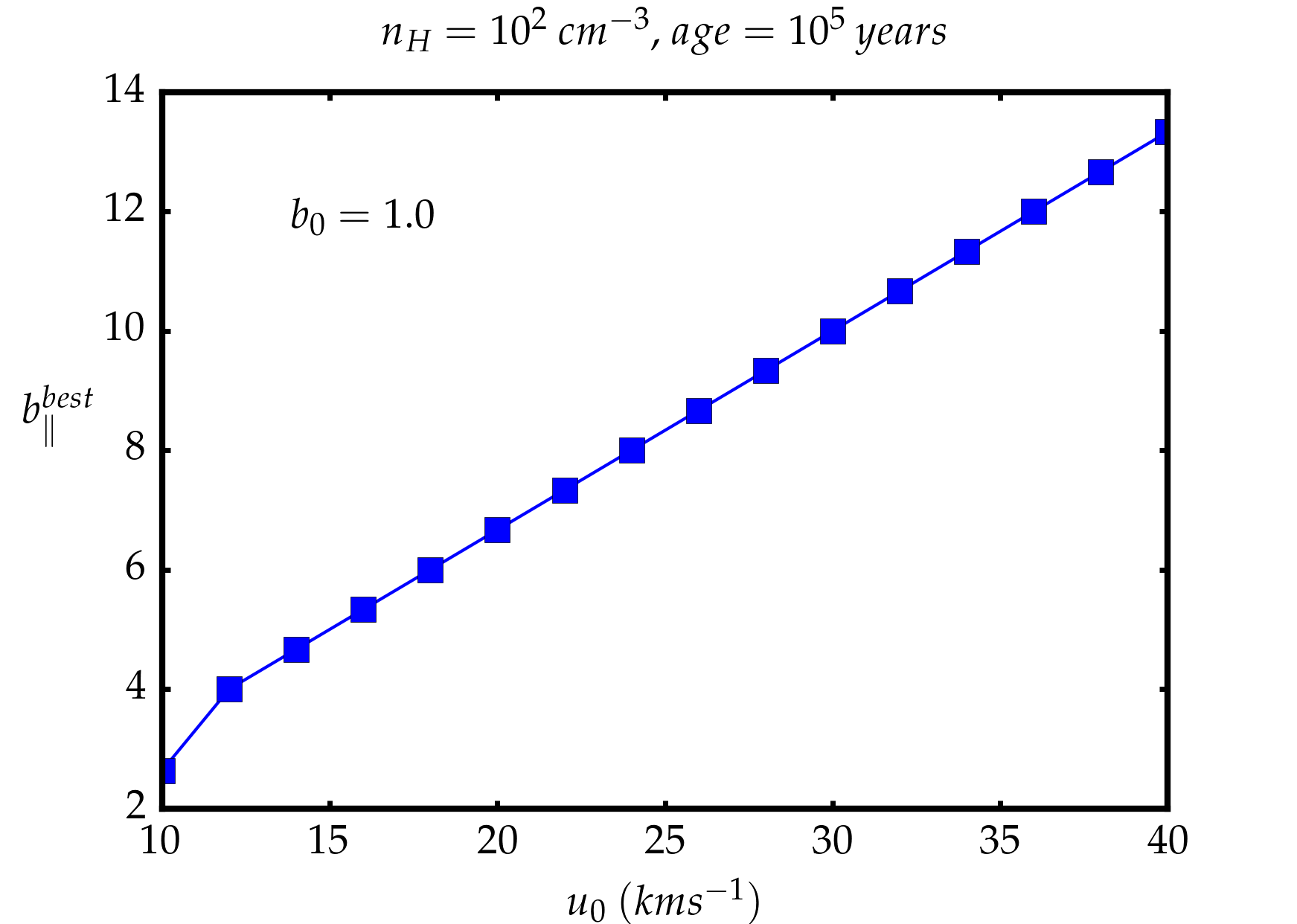}
	\end{minipage}
     \caption{Magnetization bias between 1D and 3D models. 
      The {\it top panel} is at $u_0=40$ km$\,$s$^{-1}$ and 
      for each value of $b_0$, it gives the best $b_\parallel$ 
      after the best $u_\bot$ has been determined. 
      The {\it bottom panel} is at $b_0=1$ and for 
      each value of $u_0$, it gives the best matching 
      $b_{\parallel}$ when $u_{\bot}=u_0$ is assumed. 
      Symbols are the same as in figure~\ref{fig:bias_v}. 
      The remaining parameters of the bow shock are 
      the same as for figure~\ref{fig:fit}.}
      \label{fig:bias_b}
\end{figure}

  We then turn on recovering magnetization from 1D models. We first fix the terminal velocity of the bow shock to $u_0=40$ km$\,$s$^{-1}$ and explore several values of the magnetization $b_0$, while keeping 
$\Psi=90^o$. Once the best matching 1D velocity is found, we further let the magnetization parameter $b_{\\}$ of the 1D model vary freely and explore which value best fits the 3D model (while keeping $u_\bot$ fixed). The result of this second adjustment is shown in figure~\ref{fig:bias_b}: the magnetization parameter of the best 1D model is only slightly below and represents a good match to the original magnetization parameter of the bow shock.
 Next, we assume that a priori information about the bow shock velocity (usually by looking at some molecular line width, for example) is available. We now fix $b_0=1$ for the underlying 3D model and assume  $u_\bot=u_0$ in the 1D models while searching for the best $b_\parallel$ value. The retrieved magnetization parameter is usually too high, which may lead to overesimations of the magnetization parameter when the dynamics have been constrained independently. 
  
\subsubsection{Applications and prospects}
\label{sec:app}
In this section, we briefly show how to use the 3D bow shock to interpret and constrain the parameters of bow shock observations.
\begin{figure}
	\begin{minipage}[c]{.48\textwidth}
	    \includegraphics[width=1.\linewidth, height=0.7\textwidth]
	    {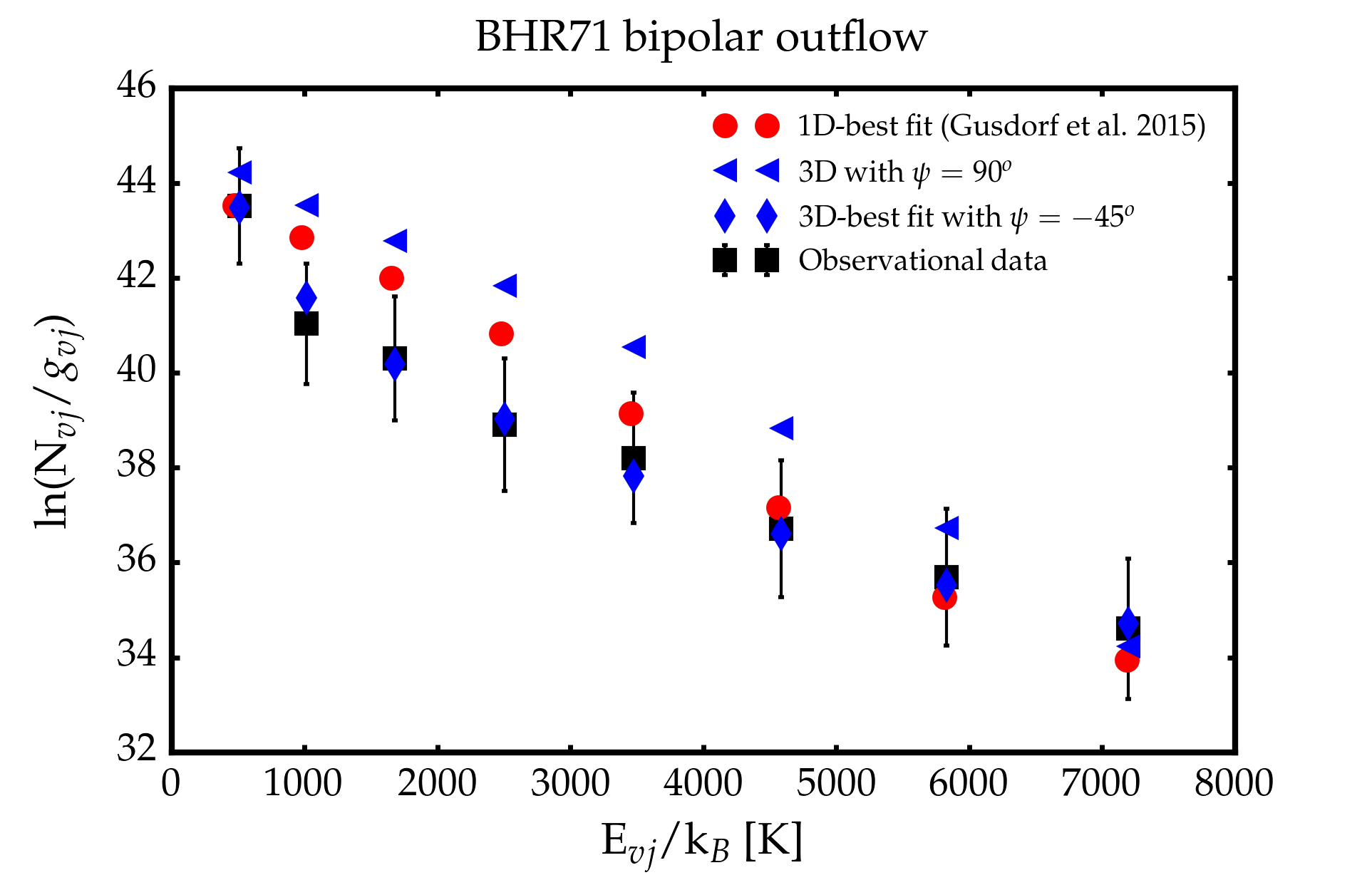}   
    	\caption{Comparison between BHR71 observations and several 
    	bow shock models. {\it Red circles}: best fit with the 1D model 
    	of \protect\cite{G15}, {\it green diamonds}: our own corresponding
     	1D model (a 3D model with velocity close to 22 km$\,$s$^{-1}$ and $\Psi=90^o$ 
     	so that the transverse magnetic field is uniform), {\it blue diamonds}:
     	best fit with our 3D model (same as the previous model, but with
     	 magnetic field orientation $\Psi=-44^o$).}
    	\label{fig:bhr71}
    \end{minipage}

	\begin{minipage}[c]{.48\textwidth}
   		\includegraphics[width=1.01\linewidth, height=0.7\textwidth]
   		{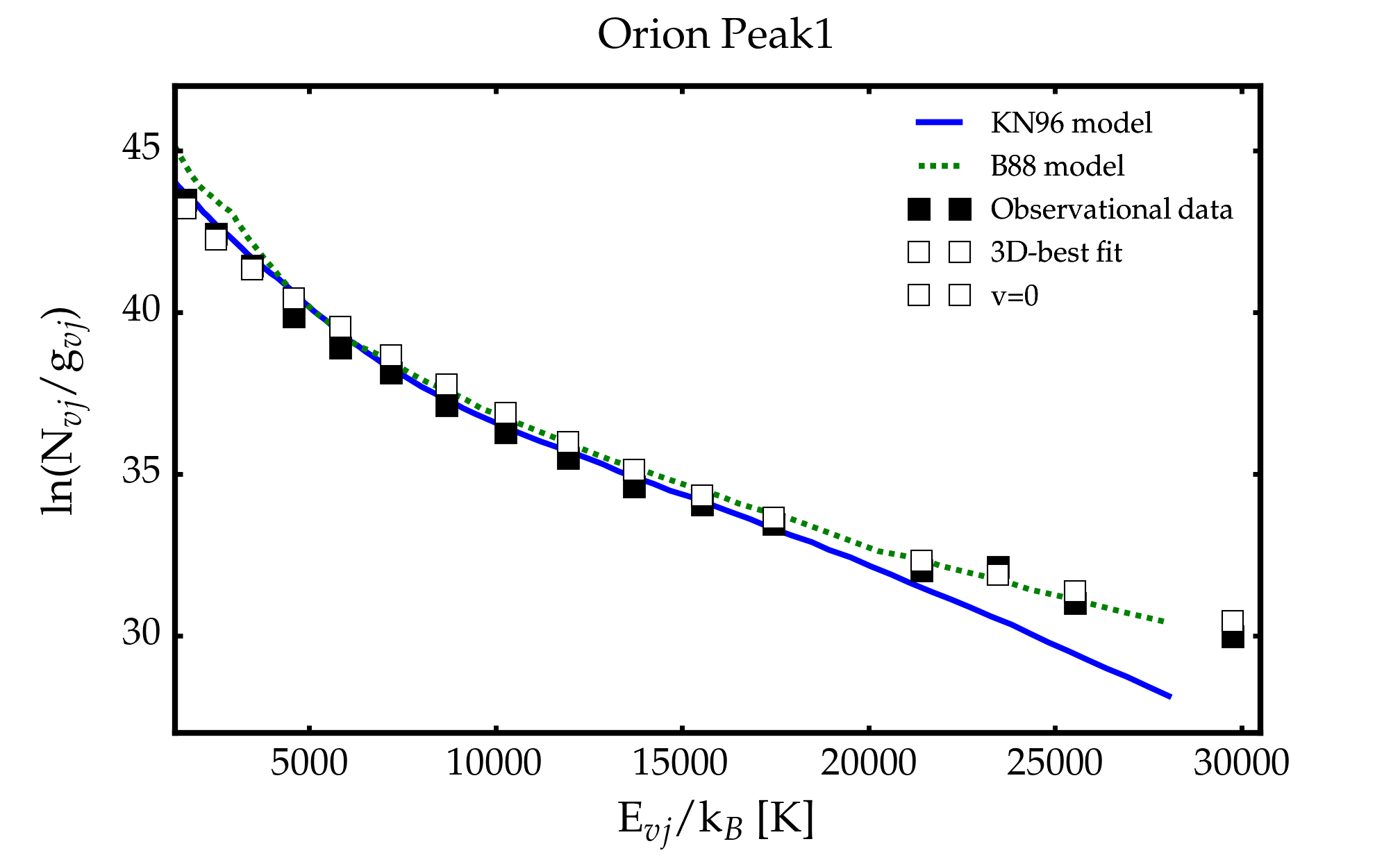}   
   		\caption{H$_{2}$ Excitation diagram observed in OMC-1 Peak1
   		 \citep{Rosenthal00} compared with various models: 
   		 our best-fit 3D-model of bow shock (open symbols), 
   		 and the best fit models from \citet{Rosenthal00}: 
   		 a combination of two planar C-shocks models from 
   		 \citet{KN96} (KN96) and one J-type shock model 
   		 from \citet{B88} (B88).}
	   \label{fig:orion_peak1}
	\end{minipage}
\end{figure}      

\subsubsection*{BHR71}  \label{subsec:BHR71_fitting}
 Located at a distance of about 175 pc (\citealt{Bourke95}), BHR71 is a double bipolar outflow (\citealt{Bourke97, Bourke01}) emerging from a Bok Globule visible in the southern sky. The two outflows are spectrally distinguishable (\citealt{Parise06}). Their driving protostars, IRS 1 and IRS 2 have luminosities of 13.5 and 0.5 L$_{\sun}$ (\citealt{Chen08}) and are separated by about 3400 AU. For this double star system, the time since collapse has been evaluated to about 36000 yr (\citealt{Yang17}). Multiple observations of this outflow system have been performed from infrared to sub-millimeter wavelength ranges. Bright HH objects HH320 and HH321 (\citealt{Corporon97}) have been detected, as well as chemical enhancement spots (\citealt{Garay98}) and several other knots of shocked gas (\citealt{Giannini04}). By combining H$_2$ observations performed by \textit{Spitzer} (\citealt{Neufeld09, Giannini11}) and SiO observations obtained from the APEX telescope, \citet{Gusdorf11} were able to characterize the non-stationary CJ-type shock waves propagating in the northern lobe of the biggest outflow. They more tightly constrained the input parameters of Paris-Durham shock models by means of successive observations of low- to higher-$J_{\rm up}$ CO (\citealt{G15}) using APEX and SOFIA. The most recent studies based on \textit{Herschel} observation hint at the presence of an atomic jet arising from the driving IRS1 protostar (\citealt{Nisini15, Benedettini17}). This does not challenge the existence of a molecular bow-shock around the so-called SiO knot position in the northern lobe of the main outflow, where most attempts have been made to compare shock models with observations (\citealt{Gusdorf11, G15, Benedettini17}). In particular, the last three studies have placed constraints on shock models of the H$_2$ emission over a beam of 24" centred on this position: pre-shock density $n_{\rm{H}} = 10^4$~cm$^{3}$, magnetic field parameter $b = 1.5$, shock velocity $v_{\rm s} = 22\ \rm km\ s^{-1}$, and age of 3800 years. In these studies, the influence of the external ISRF or from the driving protostar was neglected, with an equivalent $G_{0}$ factor set to 0. The excitation diagram that was used can be seen in figure~\ref{fig:bhr71}, where the large errorbars reflect the uncertainty on the filling factor and the proximity of the targeted region to the edge of the \textit{Spitzer}-IRS H$_2$ map. 
    
Here we attempt to reproduce the same H$_{2}$ emission data around the SiO knot position as in \cite{G15}. To fit a 3D model to this data, we should in principle adjust all the parameters in table \ref{tab:bhr71_parameters}, which would be a bit tedious, and very likely underconstrained by the observations. Instead, we started up from already published parameters and expanded around these values. We hence use a narrow range of velocities around $u_0=22$ km$\,$s$^{-1}$, $b_{0} = 1.5$ and $n_{H}=10^{4} $cm$^{-3}$ as indicated by \cite{G15}.
 These authors found an age of 3800 yr, so we took our grid models at an age of 1000 yr, as 10$^4$yr would not be compatible with the extent of the shock. A speed of 22 km$\,$s$^{-1}$ during 1000 yr already results in a shock width of 0.02 pc, about the same size of the beam (24" at 200pc according to \cite{G15}), although the H$_2$ lines emission region is a factor of a few smaller.  

  Figure~\ref{fig:bhr71} illustrates the comparison between our models and the observational values. We first restrict the velocity range in the bow shock velocity distribution to the narrow interval 
[21,23] km$\,$s$^{-1}$ that is close to the original best solution of \cite{G15}. This also accounts for the fact that the beam selects a local portion of the bow-shock and one might expect to find a privileged velocity.

 First, we examine the case $\Psi=90^o$ when the magnetization is close to $b_0$ and uniform throughout a transverse annulus of the bow-shock. Technically this is still a 3D model, but it is very close to the model in our grid of planar shocks with similar parameters because we use a very narrow range of velocities combined with uniform magnetization. The excitation diagram for this model is noted as the green diamonds in figure~\ref{fig:bhr71}. Although it slightly differs from the best model of \citet{G15}, it is not much further away from the observational constraints ($\chi=1.0$ in the model in \citet{G15} and $\chi=1.5$ in our model at $\Psi=90^o$). 

  Second, we leave the orientation of the magnetic field $\Psi$ free and we find the best model at $\Psi=-44^o$: this greatly improves the comparison with observations ($\chi=0.2$). In particular, the curvature of the excitation diagram that was difficult to model, is now almost perfectly reproduced. At this orientation, the model is a mixture of planar shocks with transverse magnetization between $b_0$ and a small minimum value. Because we limited the velocity to such a narrow range, this model is effectively a 2D model.
  
  Third, we checked that increasing the velocity range, changing the shock shape, or limiting the integration range for the angle $\varphi$ (to account for the fact that the observational beam probably intersects only one flank of the bow shock) did not improve the fit: the interpretation capabilities of our 3D model seems to be reached. Table \ref{tab:bhr71_parameters} sums up our constraints on the parameters of our model. We estimate 3-$\sigma$ error bars for $\Psi$ by investigating the shape of the $\chi^2$ well around the best value: we vary $\Psi$ with all other parameters kept fixed and we quote the range of values where $\chi^2$ is below four times its minimum value. 
  
  Finally, we checked the NY08 approximation. As mentioned in the previous section, that simple assumption works suprisingly well in the case of low pure rotational excitation. We find a best value of the power-index at $b_{SE}=2.6$, consistent with the value 2.5 in \cite{Neufeld09} for the same object, with $\chi = 0.2$: as close to the data as our 3D model.
  
\begin{table}
\begin{minipage}[c]{.45\textwidth}
\begin{tabular}{l l l}
\hline \hline 
Parameter & \hspace{1mm} Value & \hspace{1mm} Note \tabularnewline
\hline 
$n_{H}$ & $10^{4}$ cm$^{-3}$  & Pre-shock density of H nuclei \tabularnewline

$age $ & $10^{3}$ yr & Shock age \tabularnewline

$\Delta u_{\bot}$ & 21-23 km$\,$s$^{-1}$ & Range of $u_{bot}$\tabularnewline	

$b_{0}$ & 1.5 & Strengh of the magnetic field\tabularnewline

$\psi$ & $-50^{o} \pm 20^{o}$ & Orientation of the magnetic field \tabularnewline

$u_{0}$ and $\beta$ & N.A.  & Bow shock terminal velocity and \tabularnewline
$ $ & $ $ & shape are irrelevant because of \tabularnewline
$ $ & $ $ & the narrow range of velocities \tabularnewline
\hline \hline 
\end{tabular}

\caption[]{Parameters that best reproduce the excitation diagram in BHR71. We also give a $3\sigma$ uncertainty range for the parameter $\Psi$ (see text).}
\label{tab:bhr71_parameters}
\end{minipage}

\begin{minipage}[c]{.45\textwidth}
\begin{tabular}{l l l}
\hline \hline 
Parameter & \hspace{10mm} Value & \hspace{1mm} Note \tabularnewline
\hline 
$n_{H}$ & $10^{6}$ cm$^{-3}$  & Pre-shock density of H nuclei \tabularnewline

$b_{0}$ & $4.5 \pm 0.9$ & Strengh of the magnetic field\tabularnewline

$u_{0}$ & $\geq$ 30 km$\,$s$^{-1}$ & 3D terminal velocity\tabularnewline	

$age $ & $10^{3}$ yr & shock's age \tabularnewline

$\psi$ & $90^{o} \pm 30^{o}$ & Orientation of the magnetic field \tabularnewline

$\beta$ & $2.1 \pm 0.2$ & Shock shape \tabularnewline
\hline \hline 
\end{tabular}
\caption[]{Best-fit parameters of the OMC-1 Peak 1 
(see figure~\ref{fig:orion_peak1}) wit our model. 
We give an estimation of the 3$-\sigma$ uncertainty 
range for parameters $\psi$ and $b_0$ (see text).}
\label{tab:orion_parameters}
\end{minipage}
\end{table}

\subsubsection*{Orion molecular cloud}
  The Orion molecular cloud (OMC-1) is one of the well studied star forming regions. A central young stellar object generates a strong outflow that shocks the surrounding gas and yields a wealth of H$_2$ infrared emission lines that have been observed by \cite{Rosenthal00}. These authors however indicated that the full range of H$_{2}$ level population could not be reproduced by a single shock model. In fact, \cite{Bourlot02} showed that only a mixture between one J-type shock and one C-type shock model was able to account for the population of both the low and the high energy levels. 
In this work, we try to reproduce the observed excitation diagram of H$_{2}$ and strongest H$_{2}$ 1-0S(1) line profile from the OMC-1 Peak 1 with one of our bow shock models. 

  We ran a new grid of models at the pre-shock conditions in Orion, $n_{H}= 10^{6} cm^{-3}$ (\citealt{White86, B88, HM89, KN96, Kristensen08}). We limited the age to 1000 yr, which roughly corresponds to the dynamical age of the outflow (\citealt{Kristensen08}). At these densities, the shocks should have reached steady-state since long.

  Then we explore the parameter space of possible bow-shocks and seek
the best fitting model.  We considered $u_0$ between 20 and 100 km$\,$s$^{-1}$ and we varied $b_{0}$ from 1 to 6 with step 0.5. For each value of $b_{0}$,
we let the angle $\psi$ vary from $0^{o}$ to $90^{o}$ with step
$5^{o}$. Finally we explore the shape of the shock for $\beta$ in the interval from 1.0 to 3.0 with step of 0.2.

  We compute the $\chi^2$ for the 17 transitions with $v=0$ among the 55 transitions which have been measured, discarding the upper limits (table 3 of \citealt{Rosenthal00}). The parameters that best fit the excitation diagram are listed in table \ref{tab:orion_parameters}. We also provide an estimation of the 3$-\sigma$ uncertainty range for some parameters by investigating the shape of the $\chi^2$ well around the best value, as we did above for parameter $\Psi$ in the case of BHR71. The best model convincingly reproduces nearly all the lines ($\chi=0.4$), as long as the terminal velocity is greater than 30 km$\,$s$^{-1}$. The comparison to the observations is displayed in 
figure~\ref{fig:orion_peak1}: both the low and high energy regimes of the excitation diagram are obtained with the same model.  The best matching models found by \citet{Rosenthal00} are also displayed for comparison. On the other hand, they consist in a mixture of two C-type shock models from \citet{KN96} which reproduce well low energy levels, and on the other hand, in a single J-type shock model from \citet{B88}  for high energy levels. We also checked the NY08 approximation. Our best fit value is obtained at $b_{SE}=3.2$ for $\chi=0.6$. Again, this approach yields satisfying results for levels with a low excitation energy but tends to deviate at high excitation energy.  
  
\subsection{$H_{2}$ emission line profiles} \label{sec:lines}
 \cite{Smith90_H2} pioneered the study of the emission-line profile of molecular hydrogen from a simple C-type bow shock. We revisit their work using our models which improve on the treatment of shock age, charge/neutrals momentum exchange, cooling/heating functions, the coupling of chemistry to dynamics, and the time-dependent treatment of the excitation of H$_2$ molecules. We also introduce line broadening due to the thermal Doppler effect.

 In the shock's frame, the gas flows with velocity $\textbf{v}(r,u_{\bot}, \varphi) = (\hat{\textbf{t}}\ u_{\parallel} + \ \hat{\textbf{n}}\ u(r,u_{\bot},b_{\parallel})$, where $r$ is the distance within the shock thickness (orthogonal to the bow shock surface) and $u(r,u_{\bot},b_{\parallel})$ is the shock orthogonal velocity profile as computed in the 1D model. In the observer's frame, the emission velocity becomes $\textbf{v}_{\rm obs}= \textbf{v} - \textbf{u}_{0}$. However the observer only senses the component along the line of sight: $\textbf{v}_{\rm obs}.\hat{\textbf{l}}$ with $\hat{\textbf{l}}$ a unit vector on the line of sight, pointing {\it towards} the observer. When this is expressed in the observer's frame, the emission velocity becomes $v_{\rm rad}=-\textbf{v}_{\rm obs}.\hat{\textbf{l}}$.

\begin{figure*}
	\centering
	\subfloat{
		\includegraphics[height=0.35\textheight]
		{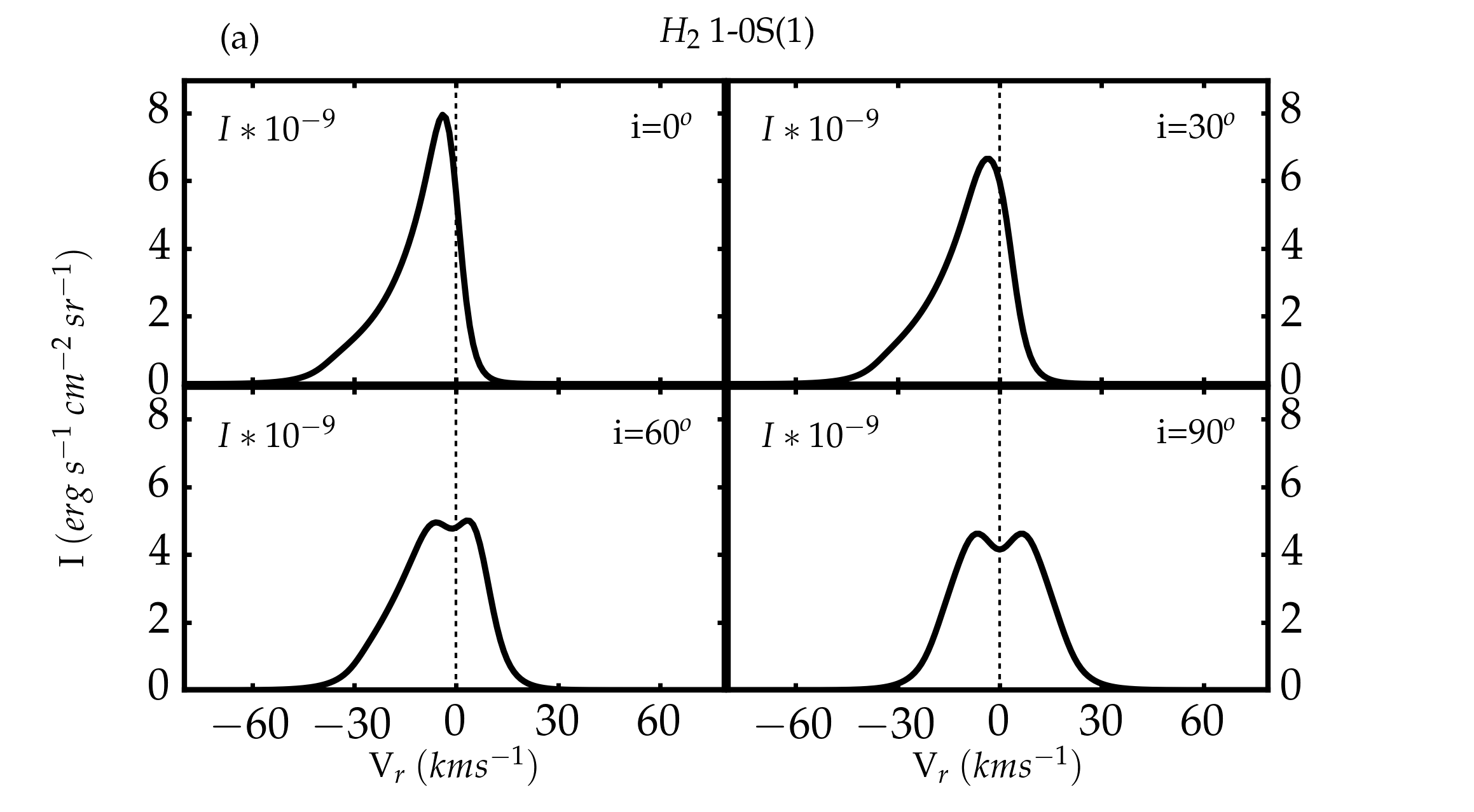}	
	}\\
	\subfloat{	
		\includegraphics[height=0.35\textheight]
		{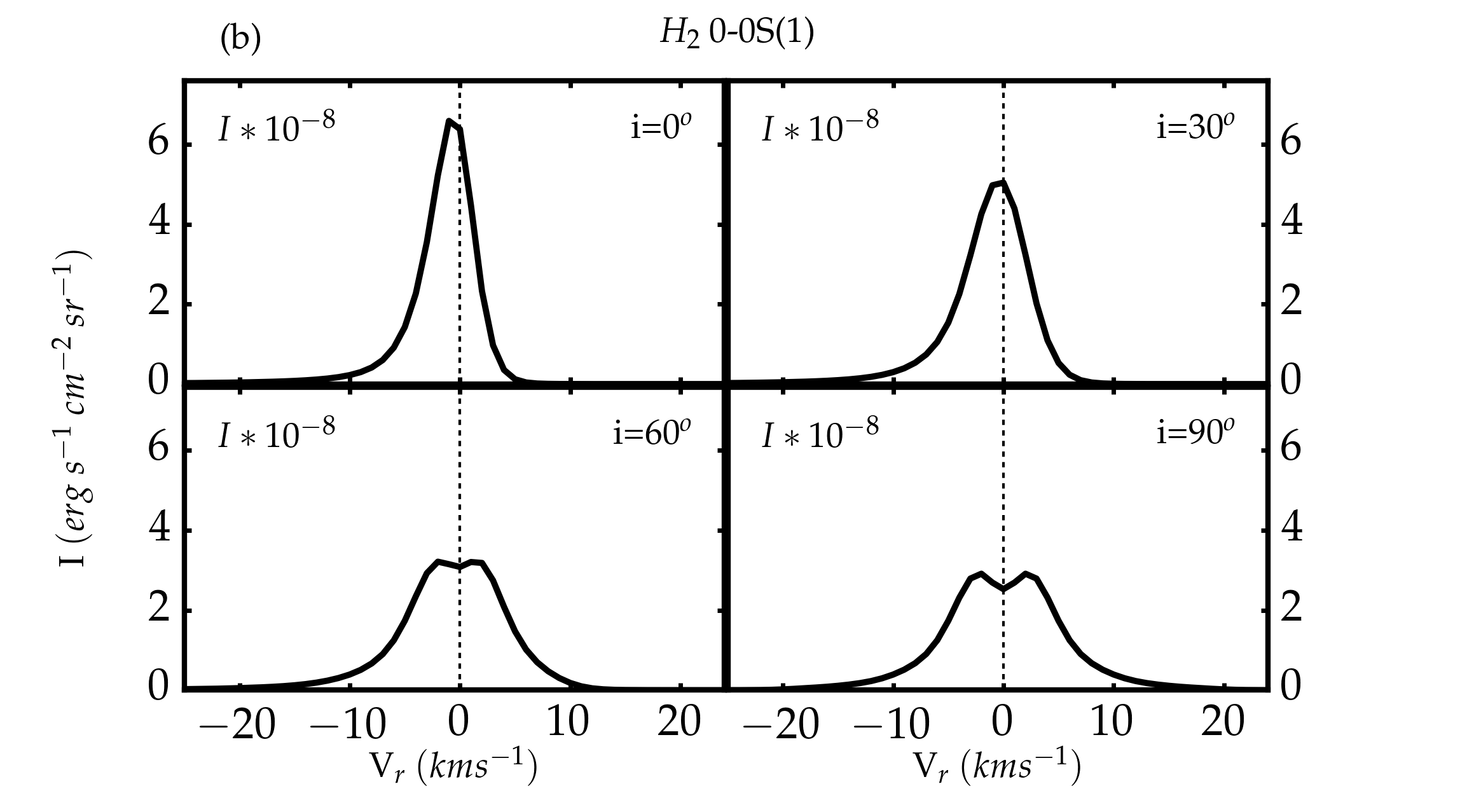}
	}
	\caption{Line profiles of a whole bow shock parameterized by 
	u$_{0}$ = 40 km$\,$s$^{-1}$, age = 10$^{2}$ years, b$_{0}$ = 1 
	and $\phi$ = 0$^{o}$. \textit{(a)} for the H$_{2}$ 1-0S(1) line and 
	\textit{(b)} for the H$_{2}$ 0-0S(1) line. 
	This figure shows the effect of the viewing angle on the line profile.}
	\label{fig:profile_vs_i}
\end{figure*}

\begin{figure*}
	\centering
	\subfloat{
	  	\includegraphics[height=0.35\textheight]
	  	{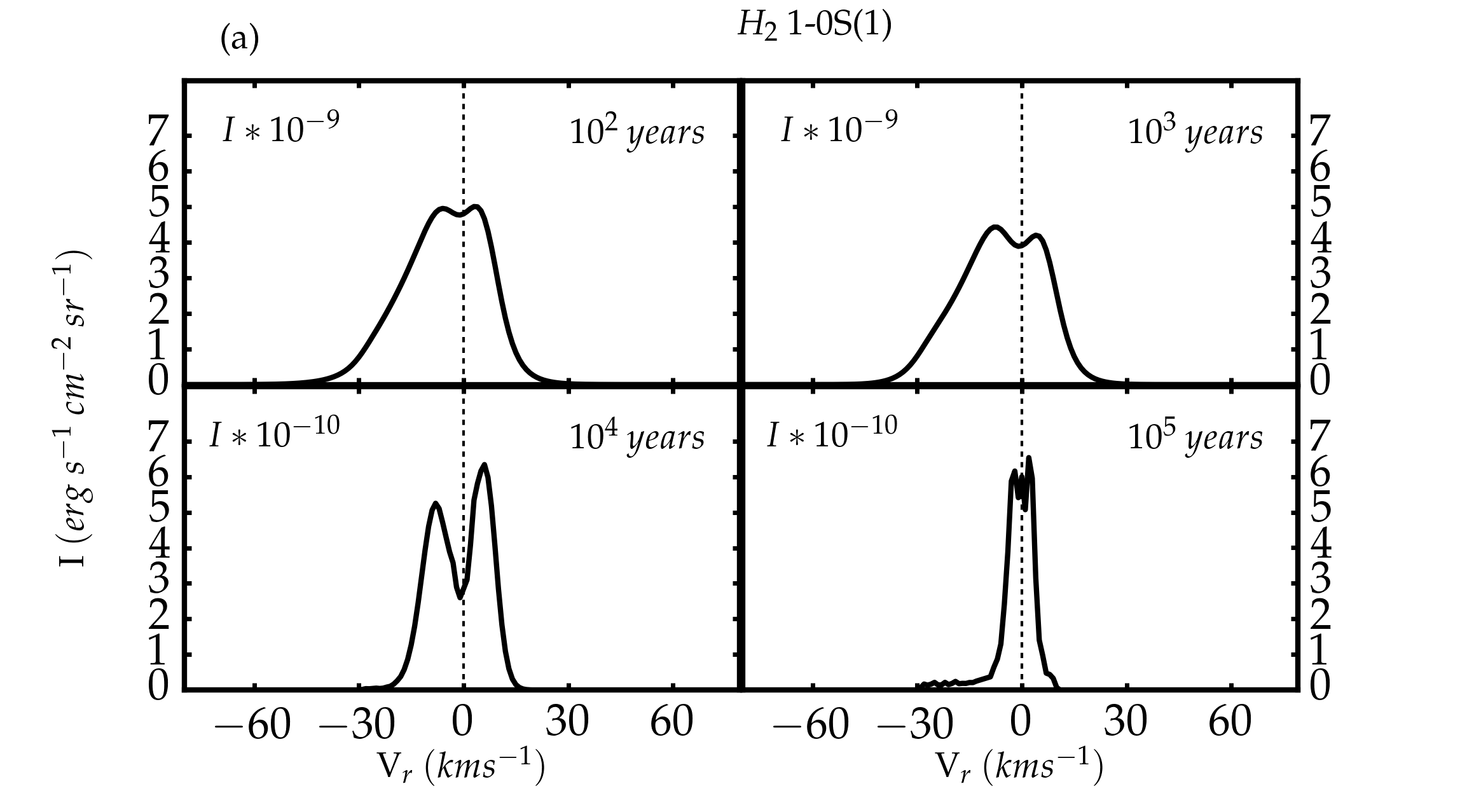}
	}\\
	\subfloat{	
 	 	\includegraphics[height=0.35\textheight]
 	 	{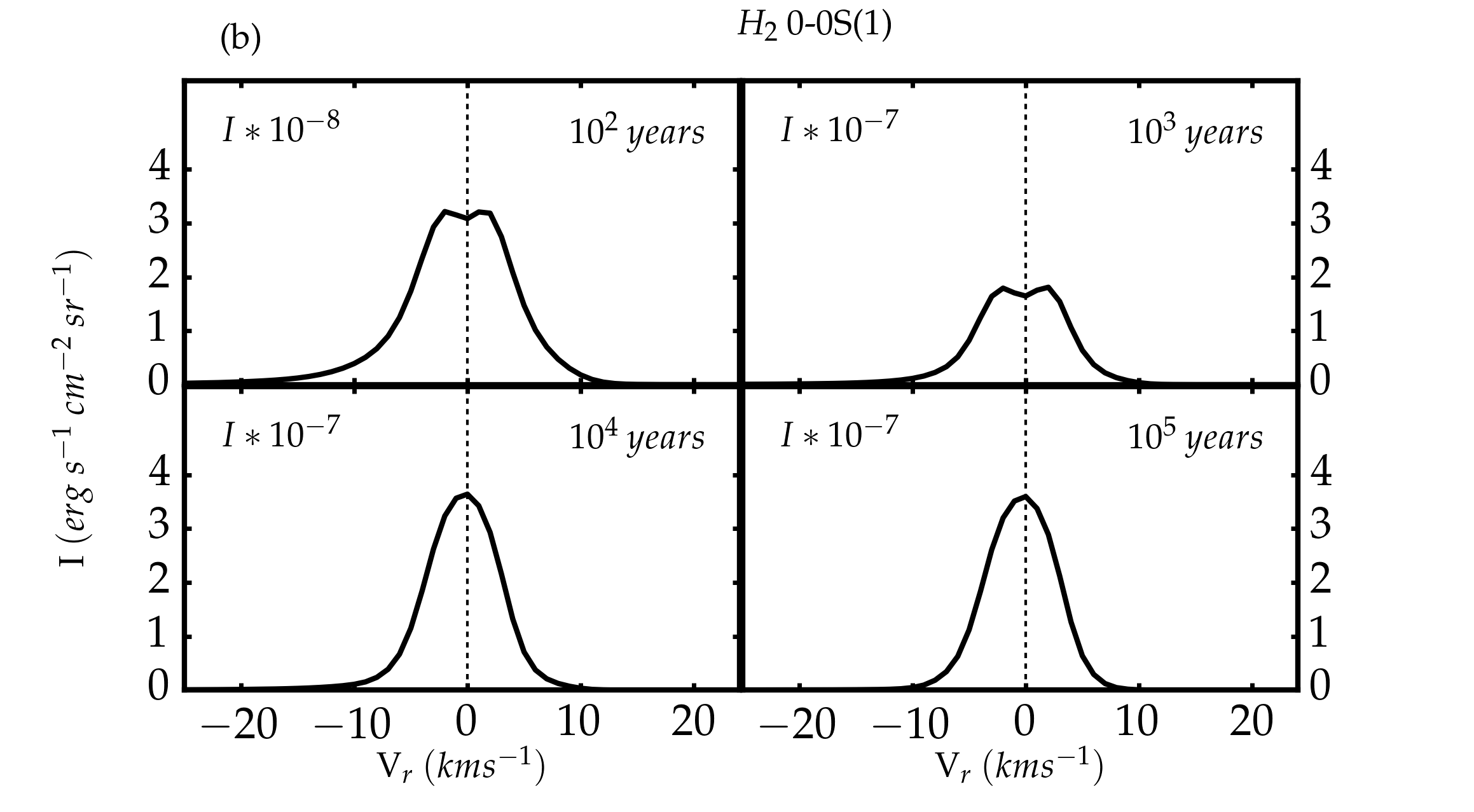}
 	}
	\caption{Line profiles of a whole bow shock parameterized by 
	u$_{0}$ = 40 km$\,$s$^{-1}$, i = 60$^{o}$, b$_{0}$ = 1 and $\phi$ = 0$^{o}$. 
	\textit{(a)} for the H$_{2}$ 1-0S(1) line and \textit{(b)} for the 
	H$_{2}$ 0-0S(1) line. This figure shows the effect of the age on the 
	line profile. Note the factor 10 change in flux scale between some panels}
	\label{fig:profile_vs_age}
\end{figure*}

 We assume the H$_2$ emission to be optically thin. Then the line profile is defined by integration over the whole volume of the bow shock, including the  emission coming from each unit volume inside each planar shock composing the bow shock. The line emission at velocity 
V$_{r}$ can be computed as follows :
	
\begin{multline}  \label{eq:line_profile}	
	f(V_r,i) = \int_{u_{\bot}}P(u_{\bot}){\rm d}u_{\bot} 
\int_{\varphi}\frac{{\rm d}\varphi}{2\pi} \\
\int_{r}{\rm d}r 
\frac{R^{2}_{0}}{\sqrt{2\pi} \sigma_{T}(r,u_{\bot},b_{\parallel})} \epsilon (r,u_{\bot},b_{\parallel}) e^{-\frac{[v_{\rm rad}(r,u_{\bot},b_{\parallel}) - V_r]^{2}}{2\sigma^{2}_{T}(r,u_{\bot},b_{\parallel})}} 
\end{multline}

which includes Doppler broadening with $\sigma^2_{T}(r,\alpha) = k_{B}/m_{H_{2}} T_{H_{2}}(r,u_{\bot},b_{\parallel})$, the thermal velocity of the H$_2$ molecule. Note that the dependence on the azimuthal angle $\varphi$ occurs both in the expression of $b_{\parallel}$  (see equation \ref{eq:b-parallel}) and the projection of $v_{\rm obs}$ onto the line-of-sight direction $\hat{\textbf{l}}$.

 Figure~\ref{fig:profile_vs_i}a shows the effect of varying the viewing angle $i$ on the 1-0S(1) line shape. When the observer looks at the bow-shock from the point of view of the star ($i=0^o$), all the emission is blue-shifted, with a stronger emission at a slightly positive velocity, coming from the part of the shock structure closest to the star, close to the J-type front where this line is excited. As the viewing angle turns more to the flank ($i$ increases), the line of sight intercepts two sides of the working surface, one going away and the other going towards the observer. The line profile then becomes doubly peaked. We checked that the integrated line emission did not vary with the viewing angle $i$.
 
Figure~\ref{fig:profile_vs_age}a shows how the age affects the 1-0S(1) line profile at a given viewing angle of 60$^o$. As the shock becomes older, the J-tail entrance velocity decreases: this explains why the two peaks of the line profile get closer to each other as age proceeds. The velocity interval between the two peaks is proportional to the entrance velocity in the J-type tail of the shocks. Furthermore, as the entrance velocity decreases, the temperature inside the J-shock decreases accordingly and the Doppler broadening follows suit: the line gets narrower as time progresses. The width of the 1-0S(1) could thus serve as an age indicator, provided that the shock velocity is well known.
  
The 0-0S(1) line corresponds to a much lower energy level than the 1-0S(1) line: while the 1-0S(1) is sensitive to temperature and shines mostly around the J-type front, the 0-0S(1) line emits in the bulk of the shock, where gas is cooler. Since the 0-0S(1) line probes a colder medium, the resulting profiles are much narrower (figure~\ref{fig:profile_vs_i}b). For early ages (100 and 1000 yr), one can however still notice the double peak signature of the J-front (figure~\ref{fig:profile_vs_age}b). Because the temperature in the magnetic precursor is much colder than the transition's upper level temperature of 1015 K for level (0,3). At these early ages, the 0-0S(1) line is shut off in the magnetic precursor (see figure~\ref{fig:CJ-temperature}, for example) and it therefore probes the J-shock part. 

\begin{figure}	
	\begin{minipage}[c]{.45\textwidth}
  		\includegraphics[width=1\linewidth]
  		{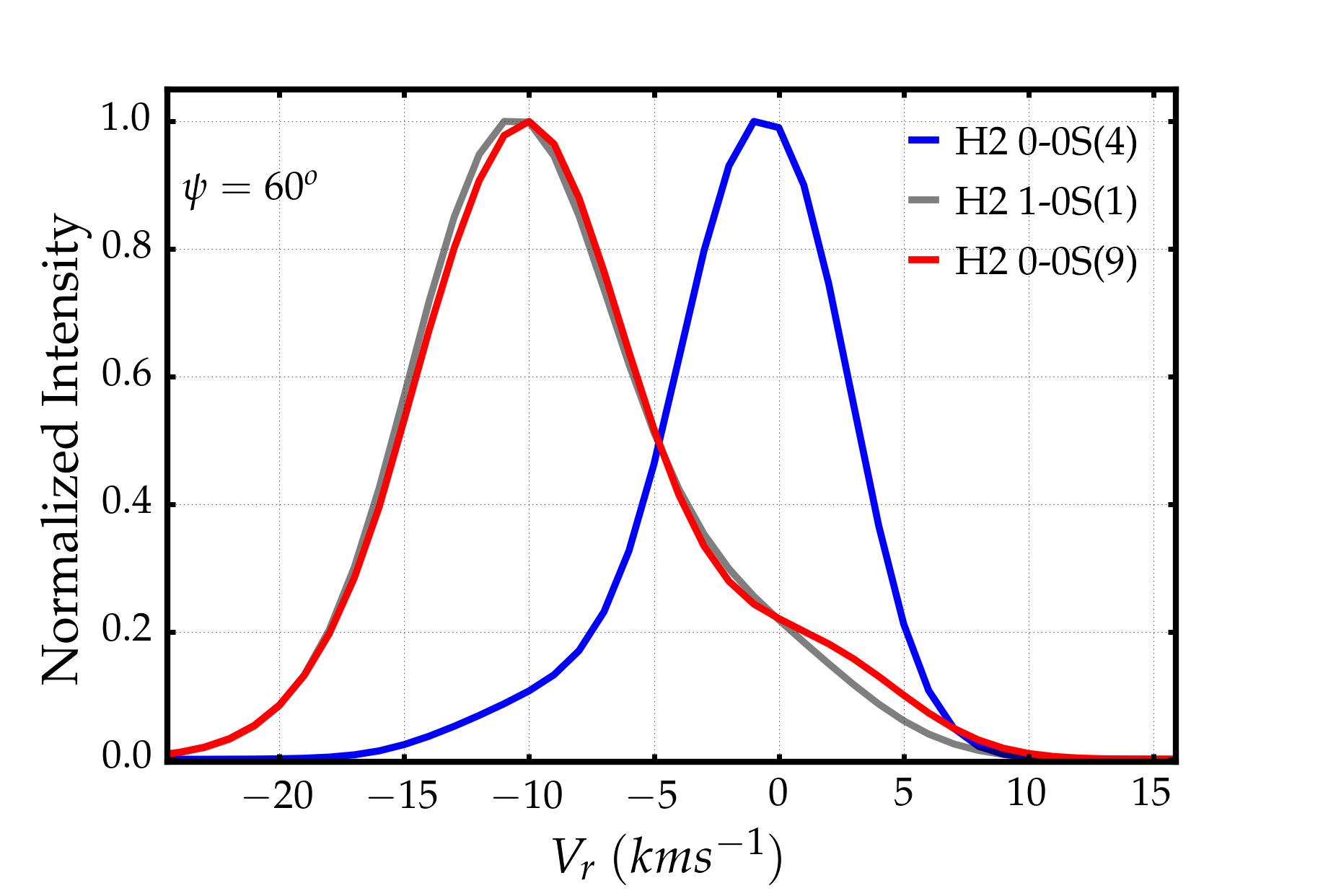}		
        \caption{Line profiles of three different transitions in a bow shock 
        at age 100 yr with parameters u$_0$=20 km$\,$s$^{-1}$, 
        n$_H$=10$^4$ cm$^{-3}$, b$_0$=1, 
        and viewing angle i = 60$^{o}$.}
        \label{fig:santangelo}
	\end{minipage}

	\begin{minipage}[c]{.45\textwidth}
  		\includegraphics[width=1\linewidth, height=0.7\textwidth]
  		{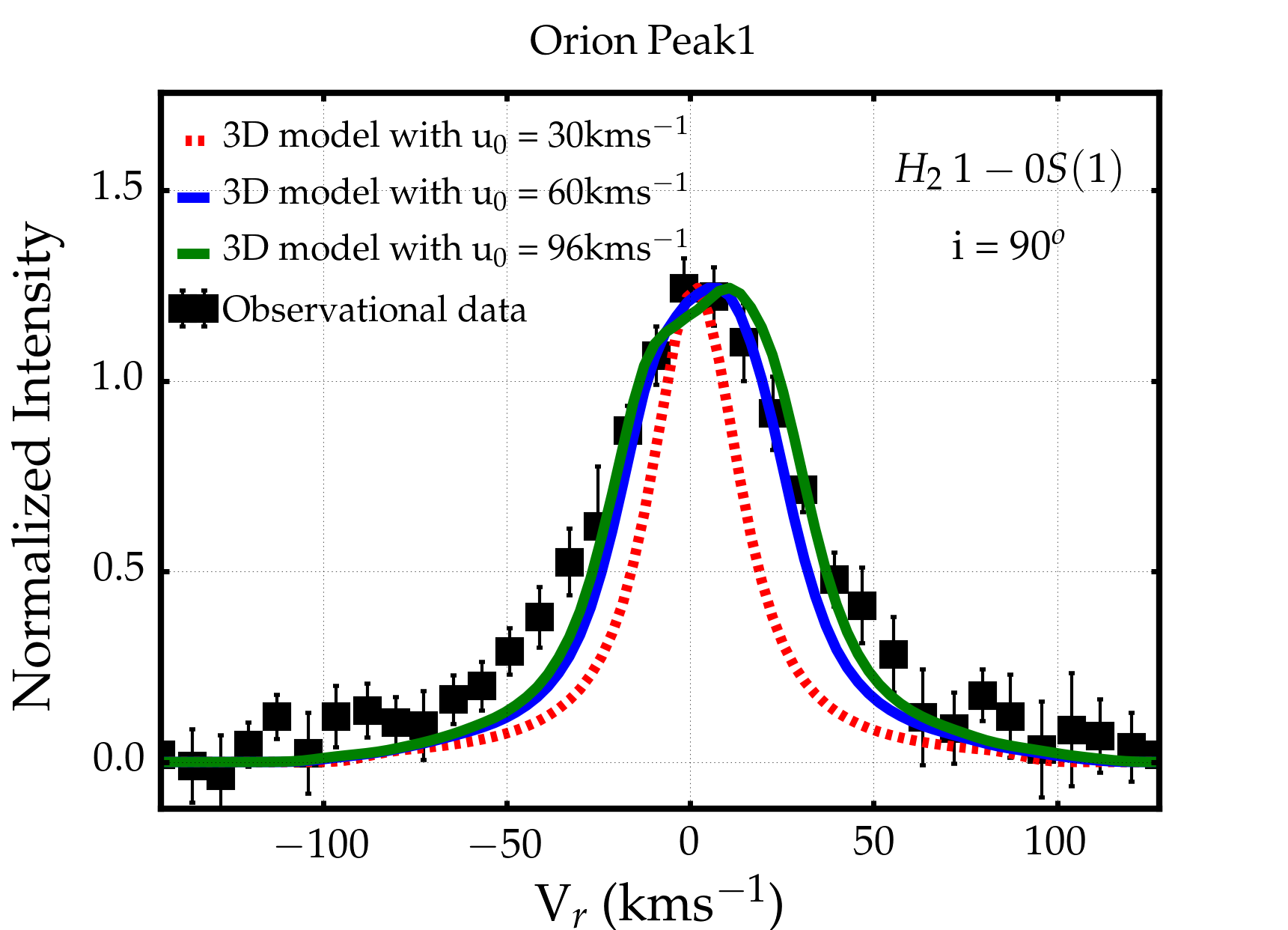}			
        \caption{Comparison of the H$_2$ line profile between OMC-1 Peak1
         observation and bow shock model. \textit{Black square:} 
         the observational data (\protect \citealt{B89}). 
         \textit{Solid lines:} our 3D model 
         using parameters in table \ref{tab:orion_parameters} with 
         different values of u$_{0}$. The best 3D model constrains 
         the terminal shock velocity to about 100 km$\,$s$^{-1}$.}
        \label{fig:Orion_profile}
	\end{minipage}

\end{figure}

 These results show that a wealth of dynamical information is contained in the line shapes. However, this information is hard to retrieve, as the line shaping process is quite convoluted. In particular, each line probes different regions of the shock depending on the upper level sensitivity to temperature. As an illustration, we plot the normalized line shapes for three different transitions in a 20 km$\,$s$^{-1}$ bow shock with pre-shock density 10$^4$ cm$^{-3}$, age 1000 yr and $b_0=1$ (figure~\ref{fig:santangelo}). This figure is meant to be compared with figure 2's top panel in \cite{San14}, which plots resolved observations of H$_2$ lines in HH54. These observations come from two different slit positions: a CRIRES slit for 1-0S(1) and 0-0S(9) near the tip of the bow, orthogonal to the outflow axis, and a VISIR slit for the 0-0S(4) line along this axis. On the other hand, our models cover the whole extent of our bow shock, which questions the validity of the comparison. Despite this, some similarities are striking: the two lines 1-0S(1) and 0-0S(9) match perfectly and are blue-shifted. The insight from our computations allows us to link the good match between the line profiles of 1-0S(1) and 0-0S(9) to the very similar energy of the upper level of the two transitions. Furthermore, we checked in our models that the emission from the low energy 0-0S(4) is completely dominated by the C-type parts of our shocks, where the velocity is still close to the ambient medium velocity: this explains why this line peaks around $V_r=0$. This C-type component should shine all over the working surface of the bow shock, and the VISIR slit along the axis probably samples it adequately. Conversely, we checked that the emission coming from both lines 1-0S(1) and 0-0S(9) is completely dominated by the J-type parts of our shocks. Hence they should shine near the tip of the bow shock (traversed by the CRIRES slit) at a velocity close to that of the star and its observed radial speed should lie around $-u_0 \cos(i)$, blue-shifted for an acute angle $i$.
          	
   \cite{B89} managed to observe a few wide H$_{2}$ line profiles from OMC-1 Peak1 by using the UKIRT telescope, configured at a  5$''$ sky aperture and with a resolution of 12 km$\,$s$^{-1}$ full width at half maximum (FWHM). A single shock model was not able to reproduce these wide observed lines (as indicated by \citealt{Rosenthal00,B89}). A C-type bow shock model of \cite{S91_Bfield} could reproduce these lines and widths, but this assumed a extremely high magnetic field strength of $\geq$ 50 mG (which amounts to b$_{\parallel}$ $\geq$ 50) while independent measurements in the same region gave much lower values: 3 mG by Zeeman splitting \citep{Norris_84} or 10mG by polarisation \citep{Chrysostomou_94}. Here we use the best parameters listed in table \ref{tab:orion_parameters} to try and reproduce the profile of the H$_{2}$ 1-0S(1) line with a more reasonable magnetisation. As mentionned in the previous subsection, the excitation diagram alone did not allow to constrain the terminal shock velocity. Now, the width of the profile allows us to constrain the velocity to about $u_0=100$ km$\,$s$^{-1}$ as illustrated by figure~\ref{fig:Orion_profile}.  The viewing angle $i\simeq$90$^{o}$ can be adjusted to the position of the peak of the line profile. Note that shock models with $u_{\bot}>$40km$\,$s$^{-1}$ are not included in these line shape models. They should contribute little to the emission since H$_2$ molecules are dissociated at high shock velocities (both due to the high temperatures experienced in these shocks and to their radiative precursors).

\section{Discussion and conclusion}
  In this study, we provide a mathematical formulation which arbitrarily links the shape of a bow shock to a distribution of planar shocks. Then, a simple convolution of this distribution with a grid of planar shocks allows to produce intensities and line shapes for any transition of the H$_2$ molecule.

  We used that property to explain the dependence of the excitation diagram of a bow shock to its parameters: terminal velocity, density, shape, age, and magnetization properties (magnitude and orientation). The combination of a steeply decreasing distribution with a threshold effect linked to the energy of the upper level of each transition yields a ``Gamow-peak'' effect. A given H$_2$ level then reaches a saturation value when the terminal velocity is above a threshold which depends directly on the energy of the level. The magnetic field  and the age dependence enter through the transition between the J-type and the C-type part of a time-dependent magnetized shock.

  The wings of a bow shock usually have a larger surface than its nose. From this, it follows that the distribution and hence the global emission properties of a bow shock are generally dominated by low-velocity shocks. A direct consequence is that the excitation diagram of a whole bow shock resembles a 1D planar shock with a lower velocity: data interpretation with 1D models is likely to be biased towards low velocity. However, if the terminal velocity of the bow shock was estimated independently (from line Doppler broadening measurements, for example), we suggest that a magnetization adjustment from 1D models to the excitation diagram will over estimate the magnetization parameter. 
Previous authors \citep[NY08,][]{Neufeld09} have suggested that the statistical equilibrium approximation could accurately reproduce observed intensities of low-energy pure rotational levels. We confirm this result, and its probable link to the distribution of entrance velocities as pointed out by NY08. However, we remark that this simple model does not satisfyingly reproduce the observations of the higher-lying transitions. A possible interpretation is that these levels are more sensitive to J-type shocks, where the sudden temperature jump is more likely to put the gas away from statistical equilibrium.

  We provide some illustrations of how our results could improve the match between model and observations in BHR71 and Orion OMC-1. We show that 3D models largely improve the interpretation. In particular, we are able to obtain much better match than in previous works with relatively little effort (and with the addition of only one or two parameters compared to the 1D models: the magnetic field orientation and the shape of the bow shock). 

  We compute line shapes with an unprecedented care and examine their dependence to age and viewing angle. Although line shapes result from a convoluted process, they contain a wealth of dynamical information. In particular, we link the double peaked structure of 1-0S(1) in young bow shocks to the dynamics of their J-type part components. The line width results from the combined effects of geometry, terminal velocity, and thermal Doppler effect. We show how different lines probe different parts of the shocks depending on the temperature sensitivity of the excitation of their upper level. We show how our 3D model can reproduce the broad velocity profile of the H$_{2}$ 1-0S(1) line in Orion Peak1 with a magnetisation compatible with other measurements. The excitation diagram fails to recover dynamical information on the velocity (it only gives a minimum value), but the line shape width provides the missing constraint.

  Further work will address some of the shortcomings of our method. First, it will be straightforward to apply similar techniques to the shocked stellar wind side of the bow shock working surface. Second, the different tangential velocities experienced on the outside and on the inner side of the working surface will very likely lead to turbulence and hence mixing, as multidimensional simulations of J-type bow shocks show. A challenge of the simplified models such as the ones presented here will be to include the mixing inside the working surface. All models presented here were run for a pre-shock ortho-para ratio of 3 : the dilute ISM is known to experience much lower ratios and we will explore the effect of this parameter on the excitation diagrams of bow-shocks in further work.
  Finally, our methods could be used to model other molecules of interest, provided that we know their excitation properties throughout the shock and that their emission remains optically thin. We expect that such developments will improve considerably the predictive and interpretative power of shock models in a number of astrophysical cases.

\section*{Acknowledgements}

We thank our referee, David Neufeld, for his careful reading of our manuscript and his enlightening suggestions. This work was mainly supported by the ANR SILAMPA (ANR-12-BS09-0025) and USTH (University of Science and Technology of Hanoi). This work was also partly supported by the French program \textit{Physique et Chimie du Milieu Interstellaire} (PCMI) funded by the Conseil National de la Recherche Scientifique (CNRS) and the Centre National d'\'Etudes Spatiales (CNES). We thank Guillaume Pineau des For{\^e}ts, Benjamin Godard and Thibaut Lebertre at LERMA/Observatoire de Paris - Paris, as well as all the members of the Department of Astrophysics (DAP) at the Vietnam National Satellite Center (VNSC) for helpful suggestions and comments. We also thank prof. Stephan Jacquemoud for a careful reading of the manuscript.



\bibliographystyle{mnras}
\bibliography{LN-TRAM_bibliotheque}

\begin{thebibliography}{}
\makeatletter
\relax
\def\mn@urlcharsother{\let\do\@makeother \do\$\do\&\do\#\do\^\do\_\do\%\do\~}
\def\mn@doi{\begingroup\mn@urlcharsother \@ifnextchar [ {\mn@doi@}
  {\mn@doi@[]}}
\def\mn@doi@[#1]#2{\def\@tempa{#1}\ifx\@tempa\@empty \href
  {http://dx.doi.org/#2} {doi:#2}\else \href {http://dx.doi.org/#2} {#1}\fi
  \endgroup}
\def\mn@eprint#1#2{\mn@eprint@#1:#2::\@nil}
\def\mn@eprint@arXiv#1{\href {http://arxiv.org/abs/#1} {{\tt arXiv:#1}}}
\def\mn@eprint@dblp#1{\href {http://dblp.uni-trier.de/rec/bibtex/#1.xml}
  {dblp:#1}}
\def\mn@eprint@#1:#2:#3:#4\@nil{\def\@tempa {#1}\def\@tempb {#2}\def\@tempc
  {#3}\ifx \@tempc \@empty \let \@tempc \@tempb \let \@tempb \@tempa \fi \ifx
  \@tempb \@empty \def\@tempb {arXiv}\fi \@ifundefined
  {mn@eprint@\@tempb}{\@tempb:\@tempc}{\expandafter \expandafter \csname
  mn@eprint@\@tempb\endcsname \expandafter{\@tempc}}}

\bibitem[\protect\citeauthoryear{{Artymowicz} \& {Clampin}}{{Artymowicz} \&
  {Clampin}}{1997}]{A97}
{Artymowicz} P.,  {Clampin} M.,  1997, \apj, \href
  {http://adsabs.harvard.edu/abs/1997ApJ...490..863A} {490, 863}

\bibitem[\protect\citeauthoryear{{Benedettini} et~al.,}{{Benedettini}
  et~al.}{2017}]{Benedettini17}
{Benedettini} M.,  et~al., 2017, \aap, \href
  {http://cdsads.u-strasbg.fr/abs/2017A\%26A...598A..14B} {598, A14}

\bibitem[\protect\citeauthoryear{{Bourke}}{{Bourke}}{2001}]{Bourke01}
{Bourke} T.~L.,  2001, \apjl, \href
  {http://cdsads.u-strasbg.fr/abs/2001ApJ...554L..91B} {554, L91}

\bibitem[\protect\citeauthoryear{{Bourke}, {Hyland}, {Robinson}, {James}  \&
  {Wright}}{{Bourke} et~al.}{1995}]{Bourke95}
{Bourke} T.~L.,  {Hyland} A.~R.,  {Robinson} G.,  {James} S.~D.,   {Wright}
  C.~M.,  1995, \mnras, \href
  {http://cdsads.u-strasbg.fr/abs/1995MNRAS.276.1067B} {276, 1067}

\bibitem[\protect\citeauthoryear{{Bourke} et~al.,}{{Bourke}
  et~al.}{1997}]{Bourke97}
{Bourke} T.~L.,  et~al., 1997, \apj, \href
  {http://cdsads.u-strasbg.fr/abs/1997ApJ...476..781B} {476, 781}

\bibitem[\protect\citeauthoryear{{Brand}, {Moorhouse}, {Burton}, {Geballe},
  {Bird}  \& {Wade}}{{Brand} et~al.}{1988}]{B88}
{Brand} P.~W.~J.~L.,  {Moorhouse} A.,  {Burton} M.~G.,  {Geballe} T.~R.,
  {Bird} M.,   {Wade} R.,  1988, \apjl, \href
  {http://adsabs.harvard.edu/abs/1988ApJ...334L.103B} {334, L103}

\bibitem[\protect\citeauthoryear{{Brand}, {Toner}, {Geballe}  \&
  {Webster}}{{Brand} et~al.}{1989}]{B89}
{Brand} P.~W.~J.~L.,  {Toner} M.~P.,  {Geballe} T.~R.,   {Webster} A.~S.,
  1989, \mnras, \href {http://adsabs.harvard.edu/abs/1989MNRAS.237.1009B} {237,
  1009}

\bibitem[\protect\citeauthoryear{{Cesarsky}, {Cox}, {Pineau des For{\^e}ts},
  {van Dishoeck}, {Boulanger}  \& {Wright}}{{Cesarsky}
  et~al.}{1999}]{Cesarsky99}
{Cesarsky} D.,  {Cox} P.,  {Pineau des For{\^e}ts} G.,  {van Dishoeck} E.~F.,
  {Boulanger} F.,   {Wright} C.~M.,  1999, \aap, \href
  {http://cdsads.u-strasbg.fr/abs/1999A\%26A...348..945C} {348, 945}

\bibitem[\protect\citeauthoryear{{Chen}, {Launhardt}, {Bourke}, {Henning}  \&
  {Barnes}}{{Chen} et~al.}{2008}]{Chen08}
{Chen} X.,  {Launhardt} R.,  {Bourke} T.~L.,  {Henning} T.,   {Barnes} P.~J.,
  2008, \apj, \href {http://cdsads.u-strasbg.fr/abs/2008ApJ...683..862C} {683,
  862}

\bibitem[\protect\citeauthoryear{{Chi{\`e}ze}, {Pineau des For{\^e}ts}  \&
  {Flower}}{{Chi{\`e}ze} et~al.}{1998}]{Ch1998}
{Chi{\`e}ze} J.-P.,  {Pineau des For{\^e}ts} G.,   {Flower} D.~R.,  1998,
  \mnras, \href {http://adsabs.harvard.edu/abs/1998MNRAS.295..672C} {295, 672}

\bibitem[\protect\citeauthoryear{{Chrysostomou}, {Hough}, {Burton}  \&
  {Tamura}}{{Chrysostomou} et~al.}{1994}]{Chrysostomou_94}
{Chrysostomou} A.,  {Hough} J.~H.,  {Burton} M.~G.,   {Tamura} M.,  1994,
  \mnras, \href {http://adsabs.harvard.edu/abs/1994MNRAS.268..325C} {268, 325}

\bibitem[\protect\citeauthoryear{{Corporon} \& {Reipurth}}{{Corporon} \&
  {Reipurth}}{1997}]{Corporon97}
{Corporon} P.,  {Reipurth} B.,  1997. p.~85

\bibitem[\protect\citeauthoryear{{DeWitt} et~al.,}{{DeWitt}
  et~al.}{2014}]{Dewitt14}
{DeWitt} C.,  et~al., 2014, European Planetary Science Congress 2014, EPSC
  Abstracts, Vol.~9, id.~EPSC2014-612, \href
  {http://cdsads.u-strasbg.fr/abs/2014EPSC....9..612D} {9, EPSC2014}

\bibitem[\protect\citeauthoryear{{Draine}}{{Draine}}{1978}]{D78}
{Draine} B.~T.,  1978, \apjs, \href
  {http://adsabs.harvard.edu/abs/1978ApJS...36..595D} {36, 595}

\bibitem[\protect\citeauthoryear{{Draine} \& {McKee}}{{Draine} \&
  {McKee}}{1993}]{Draine93}
{Draine} B.~T.,  {McKee} C.~F.,  1993, \araa, \href
  {http://adsabs.harvard.edu/abs/1993ARA\%26A..31..373D} {31, 373}

\bibitem[\protect\citeauthoryear{{Flower} \& {Pineau des For{\^e}ts}}{{Flower}
  \& {Pineau des For{\^e}ts}}{1999}]{FP99}
{Flower} D.~R.,  {Pineau des For{\^e}ts} G.,  1999, in {Ossenkopf} V.,
  {Stutzki} J.,   {Winnewisser} G.,  eds, The Physics and Chemistry of the
  Interstellar Medium.

\bibitem[\protect\citeauthoryear{{Flower} \& {Pineau des For{\^e}ts}}{{Flower}
  \& {Pineau des For{\^e}ts}}{2003}]{FP03}
{Flower} D.~R.,  {Pineau des For{\^e}ts} G.,  2003, \mnras, \href
  {http://adsabs.harvard.edu/abs/2003MNRAS.343..390F} {343, 390}

\bibitem[\protect\citeauthoryear{{Flower} \& {Pineau des For{\^e}ts}}{{Flower}
  \& {Pineau des For{\^e}ts}}{2015}]{FP15}
{Flower} D.~R.,  {Pineau des For{\^e}ts} G.,  2015, \aap, \href
  {http://adsabs.harvard.edu/abs/2015A\%26A...578A..63F} {578, A63}

\bibitem[\protect\citeauthoryear{{Flower}, {Le Bourlot}, {Pineau des
  For{\^e}ts}  \& {Cabrit}}{{Flower} et~al.}{2003}]{F03}
{Flower} D.~R.,  {Le Bourlot} J.,  {Pineau des For{\^e}ts} G.,   {Cabrit} S.,
  2003, \mnras, \href {http://adsabs.harvard.edu/abs/2003MNRAS.341...70F} {341,
  70}

\bibitem[\protect\citeauthoryear{{Garay}, {K{\"o}hnenkamp}, {Bourke},
  {Rodr{\'{\i}}guez}  \& {Lehtinen}}{{Garay} et~al.}{1998}]{Garay98}
{Garay} G.,  {K{\"o}hnenkamp} I.,  {Bourke} T.~L.,  {Rodr{\'{\i}}guez} L.~F.,
  {Lehtinen} K.~K.,  1998, \apj, \href
  {http://cdsads.u-strasbg.fr/abs/1998ApJ...509..768G} {509, 768}

\bibitem[\protect\citeauthoryear{{Giannini}, {McCoey}, {Caratti o Garatti},
  {Nisini}, {Lorenzetti}  \& {Flower}}{{Giannini} et~al.}{2004}]{Giannini04}
{Giannini} T.,  {McCoey} C.,  {Caratti o Garatti} A.,  {Nisini} B.,
  {Lorenzetti} D.,   {Flower} D.~R.,  2004, \mn@doi [\aap]
  {10.1051/0004-6361:20040087}, \href
  {http://adsabs.harvard.edu/abs/2004A26A...419..999G} {419, 999}

\bibitem[\protect\citeauthoryear{{Giannini}, {Nisini}, {Neufeld}, {Yuan},
  {Antoniucci}  \& {Gusdorf}}{{Giannini} et~al.}{2011}]{Giannini11}
{Giannini} T.,  {Nisini} B.,  {Neufeld} D.,  {Yuan} Y.,  {Antoniucci} S.,
  {Gusdorf} A.,  2011, \apj, \href
  {http://cdsads.u-strasbg.fr/abs/2011ApJ...738...80G} {738, 80}

\bibitem[\protect\citeauthoryear{{Gusdorf}, {Giannini}, {Flower}, {Parise},
  {G{\"u}sten}  \& {Kristensen}}{{Gusdorf} et~al.}{2011}]{Gusdorf11}
{Gusdorf} A.,  {Giannini} T.,  {Flower} D.~R.,  {Parise} B.,  {G{\"u}sten} R.,
   {Kristensen} L.~E.,  2011, \aap, \href
  {http://cdsads.u-strasbg.fr/abs/2011A\%26A...532A..53G} {532, A53}

\bibitem[\protect\citeauthoryear{{Gusdorf} et~al.,}{{Gusdorf}
  et~al.}{2015}]{G15}
{Gusdorf} A.,  et~al., 2015, \aap, \href
  {http://adsabs.harvard.edu/abs/2015A\%26A...575A..98G} {575, A98}

\bibitem[\protect\citeauthoryear{{Gustafsson}, {Ravkilde}, {Kristensen},
  {Cabrit}, {Field}  \& {Pineau Des For{\^e}ts}}{{Gustafsson}
  et~al.}{2010}]{Gustafsson10}
{Gustafsson} M.,  {Ravkilde} T.,  {Kristensen} L.~E.,  {Cabrit} S.,  {Field}
  D.,   {Pineau Des For{\^e}ts} G.,  2010, \aap, \href
  {http://adsabs.harvard.edu/abs/2010A\%26A...513A...5G} {513, A5}

\bibitem[\protect\citeauthoryear{{Hollenbach} \& {McKee}}{{Hollenbach} \&
  {McKee}}{1989}]{HM89}
{Hollenbach} D.,  {McKee} C.~F.,  1989, \apj, \href
  {http://adsabs.harvard.edu/abs/1989ApJ...342..306H} {342, 306}

\bibitem[\protect\citeauthoryear{{Kaufman} \& {Neufeld}}{{Kaufman} \&
  {Neufeld}}{1996}]{KN96}
{Kaufman} M.~J.,  {Neufeld} D.~A.,  1996, \apj, \href
  {http://adsabs.harvard.edu/abs/1996ApJ...456..611K} {456, 611}

\bibitem[\protect\citeauthoryear{{Kristensen}, {Ravkilde}, {Pineau Des
  For{\^e}ts}, {Cabrit}, {Field}, {Gustafsson}, {Diana}  \&
  {Lemaire}}{{Kristensen} et~al.}{2008}]{Kristensen08}
{Kristensen} L.~E.,  {Ravkilde} T.~L.,  {Pineau Des For{\^e}ts} G.,  {Cabrit}
  S.,  {Field} D.,  {Gustafsson} M.,  {Diana} S.,   {Lemaire} J.-L.,  2008,
  \aap, \href {http://adsabs.harvard.edu/abs/2008A\%26A...477..203K} {477, 203}

\bibitem[\protect\citeauthoryear{{Le Bourlot}, {Pineau des For{\^e}ts},
  {Flower}  \& {Cabrit}}{{Le Bourlot} et~al.}{2002}]{Bourlot02}
{Le Bourlot} J.,  {Pineau des For{\^e}ts} G.,  {Flower} D.~R.,   {Cabrit} S.,
  2002, \mnras, \href {http://adsabs.harvard.edu/abs/2002MNRAS.332..985L} {332,
  985}

\bibitem[\protect\citeauthoryear{{Lesaffre}, {Chi{\`e}ze}, {Cabrit}  \& {Pineau
  des For{\^e}ts}}{{Lesaffre} et~al.}{2004a}]{PL04a}
{Lesaffre} P.,  {Chi{\`e}ze} J.-P.,  {Cabrit} S.,   {Pineau des For{\^e}ts} G.,
   2004a, \aap, \href {http://adsabs.harvard.edu/abs/2004A\%26A...427..147L}
  {427, 147}

\bibitem[\protect\citeauthoryear{{Lesaffre}, {Chi{\`e}ze}, {Cabrit}  \& {Pineau
  des For{\^e}ts}}{{Lesaffre} et~al.}{2004b}]{PL04b}
{Lesaffre} P.,  {Chi{\`e}ze} J.-P.,  {Cabrit} S.,   {Pineau des For{\^e}ts} G.,
   2004b, \aap, \href {http://adsabs.harvard.edu/abs/2004A%26A...427..157L}
  {427, 157}

\bibitem[\protect\citeauthoryear{{Lesaffre}, {Pineau des For{\^e}ts}, {Godard},
  {Guillard}, {Boulanger}  \& {Falgarone}}{{Lesaffre} et~al.}{2013}]{PL13}
{Lesaffre} P.,  {Pineau des For{\^e}ts} G.,  {Godard} B.,  {Guillard} P.,
  {Boulanger} F.,   {Falgarone} E.,  2013, \aap, \href
  {http://adsabs.harvard.edu/abs/2013A\%26A...550A.106L} {550, A106}

\bibitem[\protect\citeauthoryear{{Neufeld} \& {Yuan}}{{Neufeld} \&
  {Yuan}}{2008}]{Neufeld08}
{Neufeld} D.~A.,  {Yuan} Y.,  2008, \apj, \href
  {http://adsabs.harvard.edu/abs/2008ApJ...678..974N} {678, 974}

\bibitem[\protect\citeauthoryear{{Neufeld} et~al.,}{{Neufeld}
  et~al.}{2009}]{Neufeld09}
{Neufeld} D.~A.,  et~al., 2009, \mn@doi [\apj] {10.1088/0004-637X/706/1/170},
  \href {http://adsabs.harvard.edu/abs/2009ApJ...706..170N} {706, 170}

\bibitem[\protect\citeauthoryear{{Neufeld} et~al.,}{{Neufeld}
  et~al.}{2014}]{Neufeld14}
{Neufeld} D.~A.,  et~al., 2014, \apj, \href
  {http://adsabs.harvard.edu/abs/2014ApJ...781..102N} {781, 102}

\bibitem[\protect\citeauthoryear{{Nisini} et~al.,}{{Nisini}
  et~al.}{2015}]{Nisini15}
{Nisini} B.,  et~al., 2015, \apj, \href
  {http://cdsads.u-strasbg.fr/abs/2015ApJ...801..121N} {801, 121}

\bibitem[\protect\citeauthoryear{{Norris}}{{Norris}}{1984}]{Norris_84}
{Norris} R.~P.,  1984, \mnras, \href
  {http://adsabs.harvard.edu/abs/1984MNRAS.207..127N} {207, 127}

\bibitem[\protect\citeauthoryear{{Ostriker}, {Lee}, {Stone}  \&
  {Mundy}}{{Ostriker} et~al.}{2001}]{Ostriker01}
{Ostriker} E.~C.,  {Lee} C.-F.,  {Stone} J.~M.,   {Mundy} L.~G.,  2001, \apj,
  \href {http://adsabs.harvard.edu/abs/2001ApJ...557..443O} {557, 443}

\bibitem[\protect\citeauthoryear{{Parise}, {Belloche}, {Leurini}, {Schilke},
  {Wyrowski}  \& {G{\"u}sten}}{{Parise} et~al.}{2006}]{Parise06}
{Parise} B.,  {Belloche} A.,  {Leurini} S.,  {Schilke} P.,  {Wyrowski} F.,
  {G{\"u}sten} R.,  2006, \aap, \href
  {http://cdsads.u-strasbg.fr/abs/2006A\%26A...454L..79P} {454, L79}

\bibitem[\protect\citeauthoryear{{Raga}, {de Gouveia Dal Pino},
  {Noriega-Crespo}, {Mininni}  \& {Vel{\'a}zquez}}{{Raga}
  et~al.}{2002}]{Raga02}
{Raga} A.~C.,  {de Gouveia Dal Pino} E.~M.,  {Noriega-Crespo} A.,  {Mininni}
  P.~D.,   {Vel{\'a}zquez} P.~F.,  2002, \aap, \href
  {http://adsabs.harvard.edu/abs/2002A\%26A...392..267R} {392, 267}

\bibitem[\protect\citeauthoryear{{Rosenthal}, {Bertoldi}  \&
  {Drapatz}}{{Rosenthal} et~al.}{2000}]{Rosenthal00}
{Rosenthal} D.,  {Bertoldi} F.,   {Drapatz} S.,  2000, \aap, \href
  {http://adsabs.harvard.edu/abs/2000A\%26A...356..705R} {356, 705}

\bibitem[\protect\citeauthoryear{{Santangelo} et~al.,}{{Santangelo}
  et~al.}{2014}]{San14}
{Santangelo} G.,  et~al., 2014, \aap, \href
  {http://adsabs.harvard.edu/abs/2014A\%26A...569L...8S} {569, L8}

\bibitem[\protect\citeauthoryear{{Shinn}, {Koo}, {Seon}  \& {Lee}}{{Shinn}
  et~al.}{2011}]{Shinn11}
{Shinn} J.-H.,  {Koo} B.-C.,  {Seon} K.-I.,   {Lee} H.-G.,  2011, \apj, \href
  {http://cdsads.u-strasbg.fr/abs/2011ApJ...732..124S} {732, 124}

\bibitem[\protect\citeauthoryear{{Smith}}{{Smith}}{1992}]{S92}
{Smith} M.~D.,  1992, \apj, \href
  {http://adsabs.harvard.edu/abs/1992ApJ...390..447S} {390, 447}

\bibitem[\protect\citeauthoryear{{Smith} \& {Brand}}{{Smith} \&
  {Brand}}{1990a}]{S90}
{Smith} M.~D.,  {Brand} P.~W.~J.~L.,  1990a, \mnras, \href
  {http://adsabs.harvard.edu/abs/1990MNRAS.243..498S} {243, 498}

\bibitem[\protect\citeauthoryear{{Smith} \& {Brand}}{{Smith} \&
  {Brand}}{1990b}]{Smith90_H2}
{Smith} M.~D.,  {Brand} P.~W.~J.~L.,  1990b, \mnras, \href
  {http://adsabs.harvard.edu/abs/1990MNRAS.245..108S} {245, 108}

\bibitem[\protect\citeauthoryear{{Smith}, {Brand}  \& {Moorhouse}}{{Smith}
  et~al.}{1991a}]{S91}
{Smith} M.~D.,  {Brand} P.~W.~J.~L.,   {Moorhouse} A.,  1991a, \mnras, \href
  {http://adsabs.harvard.edu/abs/1991MNRAS.248..451S} {248, 451}

\bibitem[\protect\citeauthoryear{{Smith}, {Brand}  \& {Moorhouse}}{{Smith}
  et~al.}{1991b}]{S91_Bfield}
{Smith} M.~D.,  {Brand} P.~W.~J.~L.,   {Moorhouse} A.,  1991b, \mnras, \href
  {http://adsabs.harvard.edu/abs/1991MNRAS.248..730S} {248, 730}

\bibitem[\protect\citeauthoryear{{Suttner}, {Smith}, {Yorke}  \&
  {Zinnecker}}{{Suttner} et~al.}{1997}]{Suttner97}
{Suttner} G.,  {Smith} M.~D.,  {Yorke} H.~W.,   {Zinnecker} H.,  1997, \aap,
  \href {http://adsabs.harvard.edu/abs/1997A\%26A...318..595S} {318, 595}

\bibitem[\protect\citeauthoryear{{White}, {Richardson}, {Avery}  \&
  {Lesurf}}{{White} et~al.}{1986}]{White86}
{White} G.~J.,  {Richardson} K.~J.,  {Avery} L.~W.,   {Lesurf} J.~C.~G.,  1986,
  \apj, \href {http://adsabs.harvard.edu/abs/1986ApJ...302..701W} {302, 701}

\bibitem[\protect\citeauthoryear{{Wilkin}}{{Wilkin}}{1996}]{W96}
{Wilkin} F.~P.,  1996, \apjl, \href
  {http://adsabs.harvard.edu/abs/1996ApJ...459L..31W} {459, L31}

\bibitem[\protect\citeauthoryear{{Yang}, {Evans}, {Green}, {Dunham}  \&
  {J{\o}rgensen}}{{Yang} et~al.}{2017}]{Yang17}
{Yang} Y.-L.,  {Evans} II N.~J.,  {Green} J.~D.,  {Dunham} M.~M.,
  {J{\o}rgensen} J.~K.,  2017, \apj, \href
  {http://cdsads.u-strasbg.fr/abs/2017ApJ...835..259Y} {835, 259}

\makeatother
\end{thebibliography}

\bsp	
\label{lastpage}
\end{document}